\documentclass{article}
\usepackage[margin=2cm]{geometry}
\usepackage{graphicx}

\usepackage{amsmath,amsfonts,amssymb}
\usepackage{subcaption}
\DeclareMathOperator\erf{erf}
\usepackage{color}
\usepackage{multirow}

\newcommand{\bs}[1]{\mathbf{#1}}
\usepackage{authblk}

\title{
Maximum likelihood, neural networks and posterior mean for inferring fractional Brownian motion parameters
}

\author[1]{Daniel S. Han}
\author[2]{Vaishnavi Ananthanarayanan}
\author[3,4]{Matthew J. Simpson}
\affil[1]{School of Mathematics and Statistics, University of New South Wales, Sydney, NSW, Australia}
\affil[2]{EMBL Australia Node in Single Molecule Science, Department of Molecular Medicine, School of Biomedical Sciences, University of New South Wales, Sydney, NSW, Australia}
\affil[3]{School of Mathematical Sciences, Queensland University of Technology (QUT), Brisbane, Queensland, Australia}
\affil[4]{ARC Centre of Excellence for the Mathematical Analysis of Cellular
Systems, QUT, Brisbane, Queensland, Australia}
\date{\today}

\begin{document}

\maketitle

\begin{abstract}
Fractional Brownian motion (fBM) is a model of anomalous diffusion parameterised by a Hurst exponent $H \in (0,1)$ and a generalised diffusion coefficient $D_H > 0$.
We jointly estimate $H$ and $D_H$ using maximum likelihood (MLE) from density measurements with Poisson counting noise and single-particle trajectory with Gaussian localisation error.
We show that one density measurement in time cannot infer both $H$ and $D_H$ and that several time points are needed, which matters most for inference precision.
For trajectories, we show that a neural network trained with a squared error loss approximates the posterior mean of $H$, which can be computed exactly at negligible cost.
Paired with the profile-likelihood estimate of $D_H$, the posterior mean has the lowest root mean squared error in both $H$ and $D_H$, except near $H=0$ and $H=1$ where MLE estimates $H$ better.
In vesicle trajectories from two cell types, $H$ and $D_H$ form a continuum rather than distinct modes of motion.
The posterior probability that each window within a track is subdiffusive or superdiffusive shows that $26\%$ of MRC5 fibroblast and $21\%$ of HeLa vesicle trajectories contain both subdiffusive and superdiffusive windows, against $2\%$ and $7\%$ expected from estimation uncertainty alone.
Intracellular cargoes therefore switch between subdiffusion and superdiffusion over time rather than forming populations with fixed $H$.

\end{abstract}

\section{Introduction}

There is an increasing amount of evidence for non-Brownian motion in complex systems \cite{vilk2022unravelling} ranging across intracellular transport \cite{han2020deciphering,waigh2023heterogeneous}, soft nanomaterials \cite{lyu2023electron}, financial volatility \cite{gatheral2022volatility} and foraging animals \cite{reynolds2009scale}.
The focus is placed on quantitatively characterising movement using a mathematical model, whether the object of interest is a vesicle, nanoparticle, asset price or animal, with the aim of probabilistically predicting where it will be at a future time.
Therefore, it is paramount that model parameters are accurately and precisely inferred from data.
Recent scientific studies exploiting machine learning tools have appeared on methods for detection, classification and parameter inference of anomalous diffusion models using experimental data \cite{han2020deciphering,munoz2021objective,korabel2025deep}.
A particularly useful model is fractional Brownian motion (fBM) \cite{kolmogorov1940wienersche,mandelbrot1968fractional,biagini2008stochastic} as it has a smooth one-parameter transition between sub-diffusion, diffusion and super-diffusion.
fBM's multifractional extensions, where the Hurst exponent can vary in time \cite{ayache2005multifractional,balcerek2022fractional,ayache2023uniformly}, also give a way to model heterogeneous motion.

\begin{figure}
    \includegraphics[width=\textwidth]{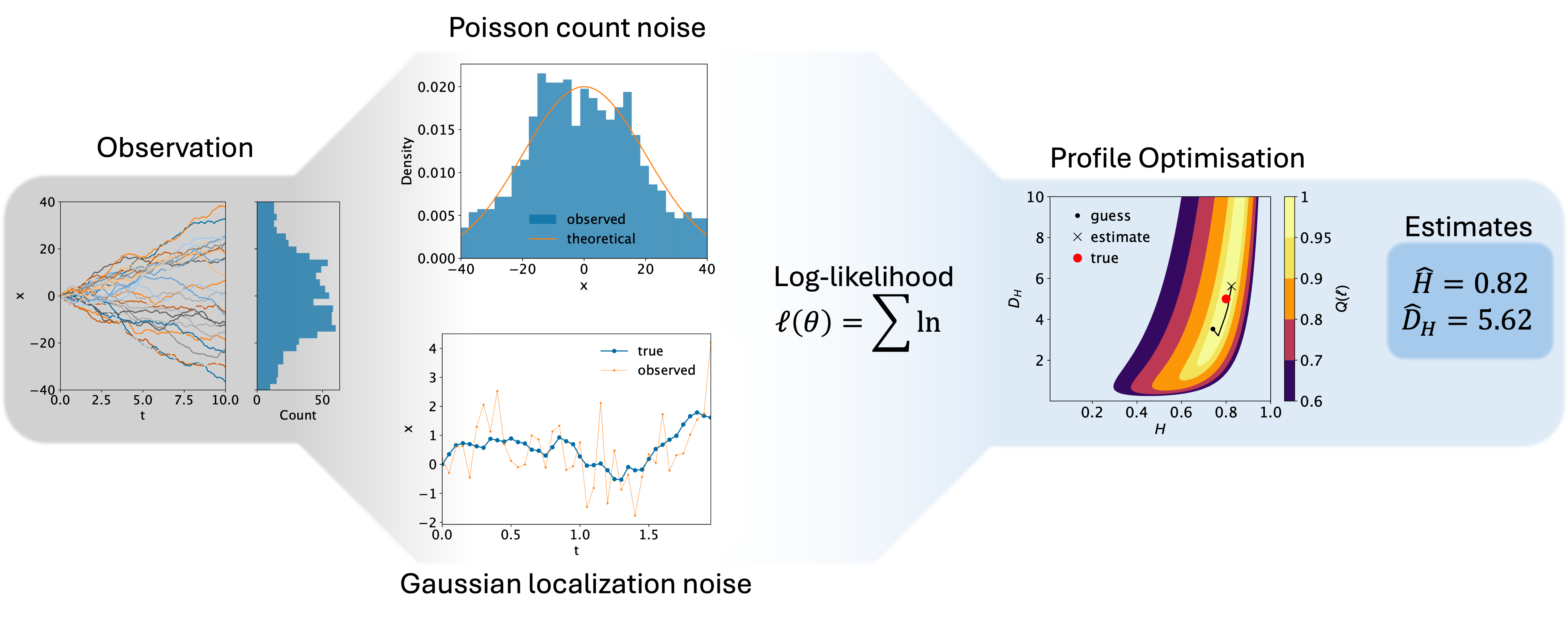}
    \caption{The parameter estimation process from observation (left) to final parameter estimates (right).}
    \label{fig:estimationPipeline}
\end{figure}

Typically, the quantitative characterisation of movement occurs at two levels of detail: from macroscopic observed density (count) measurements across a spatial domain; and from microscopic tracked paths of individual objects.
These two types of data are fundamentally different in what they reveal about the underlying process driving movement and pose different challenges for parameter inference.
A typical inference pipeline involving both levels of data is shown in Figure \ref{fig:estimationPipeline}.

Mathematically at the macroscopic level, partial differential equations (PDEs) are used to model the evolution of probability densities.
For Brownian motion (BM), we have the celebrated diffusion equation \cite{einstein1905,fick1855ueber} for which the governing PDE is
\begin{equation}
    \frac{\partial p(x,t)}{\partial t} = D \frac{\partial^2 p(x,t)}{\partial x^2},
    \label{eq:diffusionEquation}
\end{equation}
where $p(x,t)$ is the probability density function (PDF) and $D$ is the diffusion coefficient.
If the initial condition is $p(x,0)=\delta(x)$, then the  solution is
\begin{equation}
    p(x,t) = \frac{1}{\sqrt{4\pi D t}}\exp\left[-\frac{x^2}{4Dt}\right]
    \label{eq:diffusionDensity}
\end{equation}
For fBM, the governing PDE is \cite{henry2010introduction,garra2020fractional}
\begin{equation}
    \frac{\partial p(x,t)}{\partial t} = 2H t^{2H-1} D_H \frac{\partial^2 p(x,t)}{\partial x^2},
    \label{eq:fBMEquation}
\end{equation}
where the Hurst exponent $H \in (0,1)$ is a constant and $D_H > 0$ is the generalised diffusion coefficient, with solution
\begin{equation}
    p(x,t) = \frac{1}{\sqrt{4 \pi D_H t^{2H}}}\exp\left[-\frac{x^2}{4 D_H t^{2H}}\right],
    \label{eq:fBMDensity}
\end{equation}
for the same delta function initial condition.
Note that when $H=0.5$ and $D=D_H$, equations \eqref{eq:diffusionDensity} and \eqref{eq:fBMDensity} are equivalent.
Both governing PDEs \eqref{eq:diffusionEquation} and \eqref{eq:fBMEquation} can be defined on arbitrary bounded or unbounded domains.

It is important to note that the densities \eqref{eq:diffusionDensity} and \eqref{eq:fBMDensity} only specify the distribution of positions at one time such that \eqref{eq:fBMEquation} is not a Fokker-Planck equation.
These densities do not impose any conditions on how positions are correlated at different times and there is no transition density $p(x,t|x',t')$ used to derive the governing equation \eqref{eq:fBMEquation} because fBM is not a Markov process.
As a consequence, it is possible to generate the same PDE \eqref{eq:fBMEquation} and density \eqref{eq:fBMDensity} from stochastic processes that are not fBM, an example being scaled BM where the diffusion coefficient is dependent on time \cite{wang2022anomalous}.
The time-dependent coefficient $2Ht^{2H-1}D_H$ does not imply that steps grow or shrink with time, since increments of fBM are stationary, but it represents how the variance grows with time $\mathrm{Var}[x(t_0+t)-x(t_0)]=2D_H t^{2H}$.
The time $t$ in \eqref{eq:fBMDensity} is therefore the time elapsed since an arbitrary reference time $t_0$, which for fBM can be anytime along a trajectory whereas for scaled BM it is the absolute time $t_0=0$.
Usually, the arbitrary reference time $t_0$ is when the experiment or observation begins.
Every statistic or likelihood that involves multiple time points for fBM must therefore be viewed through its covariance (explored in Section \ref{sec:MLEtrajectories}) rather than through \eqref{eq:fBMDensity}.
As a result, the mean density cannot distinguish between processes that share the same one-time distributions because the density based likelihoods treat multiple time observations as independent samples.
This implies that inference problems based on only density data are inherently ill-posed \cite{sposini2022towards}, unless one can prove a unique mapping between the PDE and the underlying stochastic process.

The joint distribution of positions across time points is required to resolve this ambiguity, which is exactly what the microscopic level provides.
The PDEs \eqref{eq:diffusionEquation} and \eqref{eq:fBMEquation} can be obtained from their corresponding stochastic differential equations (SDEs).
For BM, the SDE is
\begin{equation}
    dX(t) = \sqrt{2D}dB(t),
\end{equation}
where $X(t)$ is the position of a single trajectory at time $t$ and $B(t)$ is a Wiener process with expectation $\mathbb{E}\left[B(t)\right]=0$ and covariance
\begin{equation}
\mathbb{E}\left[B(t)B(s)\right] = \min(t,s).
    \label{eq:bm_covariance}
\end{equation}
Here, the expectation, $\mathbb{E}[\cdot]$, is the average over independent realisations of the process, i.e.\ an ensemble average over trajectories at fixed times \cite{oksendal2003stochastic,karatzas1991brownian}.
Since $\mathbb{E}[B(t)]=0$, the expectation in \eqref{eq:bm_covariance} is the covariance,
\begin{equation*}
    \mathrm{Cov}\left[B(t),B(s)\right] = \mathbb{E}\left[B(t)B(s)\right]-\mathbb{E}\left[B(t)\right]\mathbb{E}\left[B(s)\right] = \mathbb{E}\left[B(t)B(s)\right],
\end{equation*}
which measures how strongly the position at time $t$ varies together with the position at time $s$ across realisations.
Setting $s=t$ in \eqref{eq:bm_covariance} gives the variance $\mathbb{E}[B(t)^2]=t$, so the MSD of BM is $\mathbb{E}[X(t)^2]=2Dt$.
For fBM the corresponding representation (not an It\^o SDE) is
\begin{equation}
    X(t) = \sqrt{2 D_H}\, B_H(t),
    \label{eq:fbm_SDE_d}
\end{equation}
where $B_H(t)$ is a fractional Brownian motion \cite{mandelbrot1968fractional,samorodnitsky1994stable,biagini2008stochastic}, with mean $\mathbb{E}\left[B_H(t)\right]=0$ and covariance
\begin{equation}
    \mathbb{E}\left[B_H(t)B_H(s)\right] = \frac{1}{2}\left( t^{2H}+s^{2H}-|t-s|^{2H} \right).
    \label{eq:fbm_process_covariance}
\end{equation}
Similar to BM, \eqref{eq:fbm_process_covariance} is the covariance of $B_H$ because its mean is zero, and setting $s=t$ gives $\mathbb{E}[B_H(t)^2]=t^{2H}$, so $\mathbb{E}[X(t)^2]=2D_H t^{2H}$.
The distinction between BM and fBM lies in the covariance as the increments for BM are independent but for fBM have long-range correlation.
These correlations that can only be measured through trajectory level data identify whether the underlying stochastic process for an experimental system is fBM or scaled BM.

Traditionally, trajectory data are characterised via the mean-squared displacement (MSD), $\langle x^2 (t) \rangle = 2D_H t^{2H}$ \cite{henry2010introduction}, which is fitted to estimate $H$ and $D_H$.
The estimated exponent classifies a trajectory as subdiffusive ($H<1/2$), diffusive ($H=1/2$) or superdiffusive ($H>1/2$).
This classification assigns one type of motion to an entire trajectory.
Estimation of $H$ from short sections of the trajectory (windows) is needed to distinguish if a single trajectory switches between subdiffusion and superdiffusion over time.
Depending on the accuracy of this estimation, every classification also needs uncertainty quantification to gauge the confidence of trajectory segment classification.
However, MSD fitting is known to be statistically inefficient and unreliable, particularly for short and noisy data \cite{qian1991single,berglund2010statistics,kepten2013improved,vestergaard2014optimal,kepten2015guidelines}.
This problem has driven multiple alternative approaches, notably measuring autocorrelation \cite{fulinski2017fractional,grzesiek2024distinguishing} or coefficient of variation \cite{sposini2022towards} and using machine learning, in particular deep neural networks \cite{han2020deciphering,munoz2021objective,korabel2025deep}.
While neural networks are highly accurate, training for parameter estimation does not return a measure of uncertainty on an estimate.
The Bayesian deep-learning variants \cite{seckler2022bayesian} that do quantify uncertainty calibrate uncertainty using the distribution drawn from the training set rather than against the likelihood of the data.
In both cases, it is difficult to know when an estimate can be trusted and its uncertainty.

In this work, we take a likelihood-based approach to parameter inference for fBM shown in Figure \ref{fig:estimationPipeline}.
This is attractive because it is principled and it yields both the maximum likelihood estimates of $H$ and $D_H$ as well as their uncertainty and identifiability \cite{Pawitan2001,Cole2020,Simpson2023,Villaverde2023}.
We consider both levels of data: density count measurements subject to Poisson noise; and individual trajectories subject to Gaussian noise.
Many estimators target $H$ alone and leave $D_H$ as a scale factor to be determined afterwards, but since the fBM likelihood depends on $H$ and $D_H$ jointly, we estimate and assess them together in this work.
We use the profile likelihood to quantify how data quantity and quality affects the precision of $H$ estimation.
This quantification directly measures the ability to distinguish between BM and fBM through parameter estimation.
Likelihood-based inference for fBM is well established \cite{lundahl1986fractional,harba1992fractional,dai2021maximum} and known to be asymptotically efficient \cite{dahlhaus1989efficient}.
Recently, fBM trajectory likelihoods accounting for localisation error and realistic experimental conditions have also been developed \cite{backlund2015chromosomal,koo2016systems,mellnik2016maximum,lanoiselee2018optimal}, with Bayesian extensions \cite{krog2018bayesian,thapa2018bayesian}.
For density level data, likelihood-based inference is also established \cite{liu2024likelihood,shi2025fractional}.
The contribution of this work is a systematic assessment of what likelihood estimation can and cannot deliver with density and trajectory level data.
For density data, we quantify how the number of observation times, the gain and the background noise affect the accuracy and precision of inferring $H$ and $D_H$.
For trajectory data, we quantify how the localisation noise and number of time points affect the inference accuracy and precision.
We also demonstrate and explain the mechanisms behind pathological cases when the trajectory MLE can rail to the boundaries $H=0$ and $H=1$.

Furthermore, we compare the MLE with a hybrid neural network method \cite{han2020deciphering}, the exact posterior mean of $H$ and various commonly used summary statistics, and find that pairing the posterior mean of $H$ with the profile-likelihood estimate of $D_H$ (PM+pMLE) has a lower root mean squared error (RMSE) than MLE except near $H=0$ and $H=1$.
We then apply these methods to infer $H$ and $D_H$ from intracellular vesicle trajectories from two different existing datasets \cite{han2020deciphering,tirumala2024single}.
Using the posterior probability that each window of a trajectory is subdiffusive, we show that $21-26\%$ of trajectories contain both subdiffusive and superdiffusive windows, far more than expected from chance and estimation uncertainty alone ($2-7\%$).
This shows that intracellular cargoes switch between subdiffusion and superdiffusion over time rather than forming populations with a fixed $H$, motivating models where $H$ varies over time.

\section{MLE for Densities}
\label{sec:MLEdensities}

\begin{figure}
    \centering
    \includegraphics[width=\linewidth]{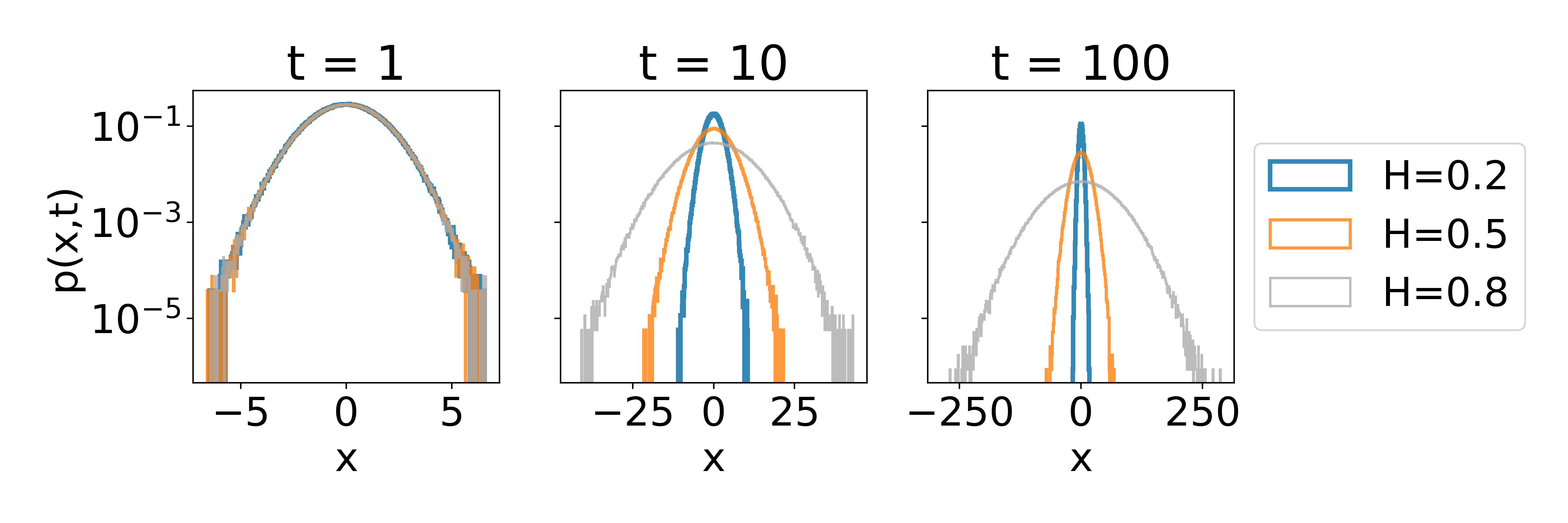}
    \caption{Histograms showing the densities of subdiffusive fBM ($H=0.2$), diffusion ($H=0.5$) and superdiffusive fBM ($H=0.8$) at time points $t=1$, $10$ and $100$ with $D_H = 1$.}
    \label{fig:densities}
\end{figure}

Experimentally, it is often convenient to observe overall densities from images acquired of a natural system.
The advantage of measuring densities as opposed to single particle tracking is that it is computationally less expensive and removes the dependence of data quality on the tracking algorithm or acquisition noise.
For example, fluorescent tags are commonly used in biology to mark objects of interest and to characterise the dynamics of the movement of lipids, proteins and cytosolic fluid from one cellular compartment to another, as well as the motion of tagged individual cells within complex populations of heterogeneous cells \cite{Simpson2013,Haridas2017}.
These measurements are strictly non-negative observations that influence our modelling choices below.
For this reason, we outline how MLE of fBM model parameters can be applied to an observed density and explore the accuracy and reliability of estimates with different experimental parameters such as the number of time points and magnitude of measurement noise.
Figure~\ref{fig:densities} shows the PDF of particles performing BM and fBM, with initial condition $p(x,0)=\delta(x)$, for different times, $t$, and different Hurst exponents, $H$.

\begin{figure}
    \centering
    \includegraphics[width=0.5\linewidth]{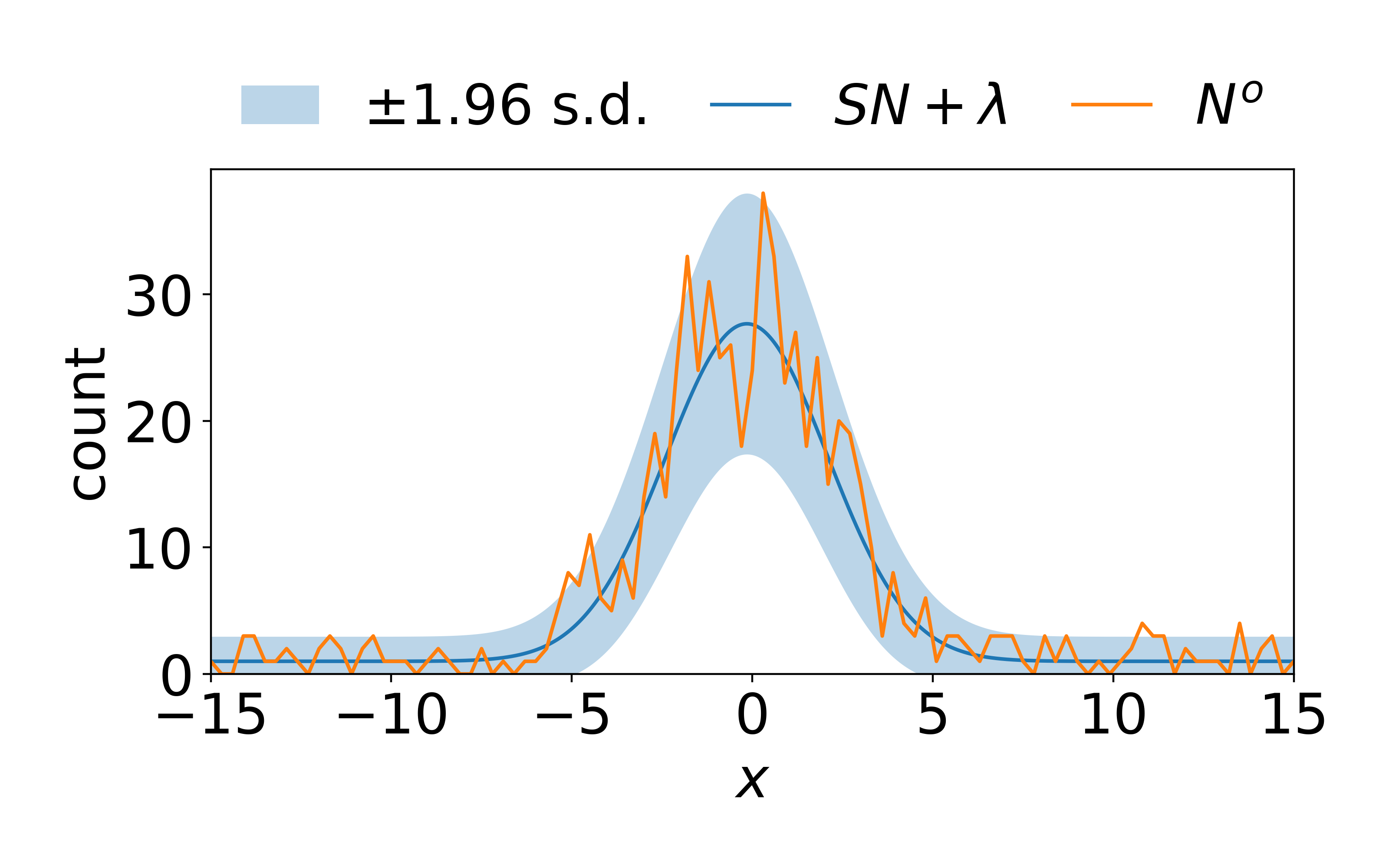}
    \caption{Plots at $t=10$ of the mean observed count $SN(x,t)+\lambda$ (blue line) with $\pm1.96$ standard deviations of the Poisson noise model \eqref{eq:PoissonNoiseModel} (blue shaded), where $N(x,t)$ is given by \eqref{eq:fBM_counts_exact} with $D_H=1$ and $H=0.2$, and one realisation of $N^o(x,t)$ (orange line), with $S=1$ and $\lambda =1$.}
    \label{fig:PoissonNoisePDF}
\end{figure}

Consider the discrete problem of observing a count, $N^o(x_i,t)$, in a bin of width $\Delta x$, given a model that predicts the mean to be $N(x_i,t)$.
To relate this count to a PDF, one would normalize the histogram as if it were a density accounting for bin widths.
We assume the model has Poisson noise such that the observed count, $N^o(x_i,t) \sim \text{Pois}(S\cdot N(x_i,t)+\lambda)$. The probability of observing $N^o(x_i,t)$ counts for the bin interval $[x_i,x_i+\Delta x]$ is
\begin{equation}
\rho(\theta, N^o;N) \equiv \text{Prob}\{N^o|N\} = \frac{(S\cdot N+\lambda)^{N^o} e^{-(S\cdot N+\lambda)}}{N^o!},
\label{eq:PoissonNoiseModel}
\end{equation}
where we have dropped the arguments in $N$ and $N^o$ above for readability. In this case, $\theta = (\theta_{PDF},S,\lambda)$, where $\theta_{PDF} = (D_H,H)$ are the parameters related to the model PDF.
The additive term $\lambda \geq 0$ is the mean background count per bin, which in an experiment corresponds to dark current, autofluorescence and stray light.
The multiplicative term $S>0$ is the effective gain, the product of collection efficiency, quantum efficiency and camera gain, so that $SN$ is the mean number of detected counts for the objects of interest.
In an ideal world, the perfect measurement device would have $S=1$ and $\lambda =0$.
The random quantity in this model is the observed count, $N^o(\cdot,t)$, which is an independent Poisson random variable at every observation time.
This model replicates experiments that acquire a fluorescence image, where the detected counts are shot noise about a smooth intensity or sample population.
Figure~\ref{fig:PoissonNoisePDF} shows a realisation of $N^o(x,t)$ given the fBM model, $N(x,t)$.
Visually, the variance in observation $N^o(x,t)$ is largest at $x=0$, corresponding to the largest values of $N(x,t)$.

According to \eqref{eq:PoissonNoiseModel}, we can write the likelihood of observing the counts in $M$ bins, $[x_1,x_1+\Delta x]$, $[x_2,x_2+\Delta x]$, ..., $[x_M,x_M+\Delta x]$, as
\begin{equation}
    L(\theta,N|N^o) = \prod_{i=1}^{M}\rho(\theta,N^o(x_i,t);N(x_i,t)),
    \label{eq:likelihood_poisson}
\end{equation}
and the loglikelihood as
\begin{equation}
    \ell(\theta,N|N^o) = \sum_{i=1}^{M}\ln(\rho(\theta,N^o(x_i,t);N(x_i,t))).
    \label{eq:loglikelihood_poisson}
\end{equation}
Note that these likelihoods are only for single time observations and the generalisation to multiple time points is straightforward as we will show in Section \ref{sec:nonidentifiability_densities}.
Given a total particle count $\nu$, $N$ and $N^o$ can be related back to the PDF in \eqref{eq:fBMDensity}, $p(x,t)$, which is the solution to the PDE \eqref{eq:fBMEquation} through \begin{equation}
\begin{split}
N(x,t) &= \nu \int_{x}^{x+\Delta x}p(y,t)dy\\
&= \frac{\nu}{2}\left( \erf\left(\frac{x+\Delta x}{\sqrt{4 D_H t^{2H}}}\right) - \erf\left(\frac{x}{\sqrt{4 D_H t^{2H}}}\right) \right)
\end{split}
\label{eq:fBM_counts_exact}
\end{equation}
or the approximation
\begin{equation}
\begin{split}
N(x,t) &= \nu p(x,t)\Delta x\\
&= \frac{\nu\Delta x}{\sqrt{4\pi D_H t^{2H}}}\exp\left[-\frac{x^2}{4D_Ht^{2H}}\right],
\end{split}
\end{equation}
for small $\Delta x$. This is usually the case in experiments and $\Delta x$ is also usually a known constant related to the pixel resolution of the acquisition method. The simulation and fits that follow use \eqref{eq:fBM_counts_exact}.

\subsection{Non-identifiability and experimental design}
\label{sec:nonidentifiability_densities}

\begin{figure}
    \centering
    \includegraphics[width=\linewidth]{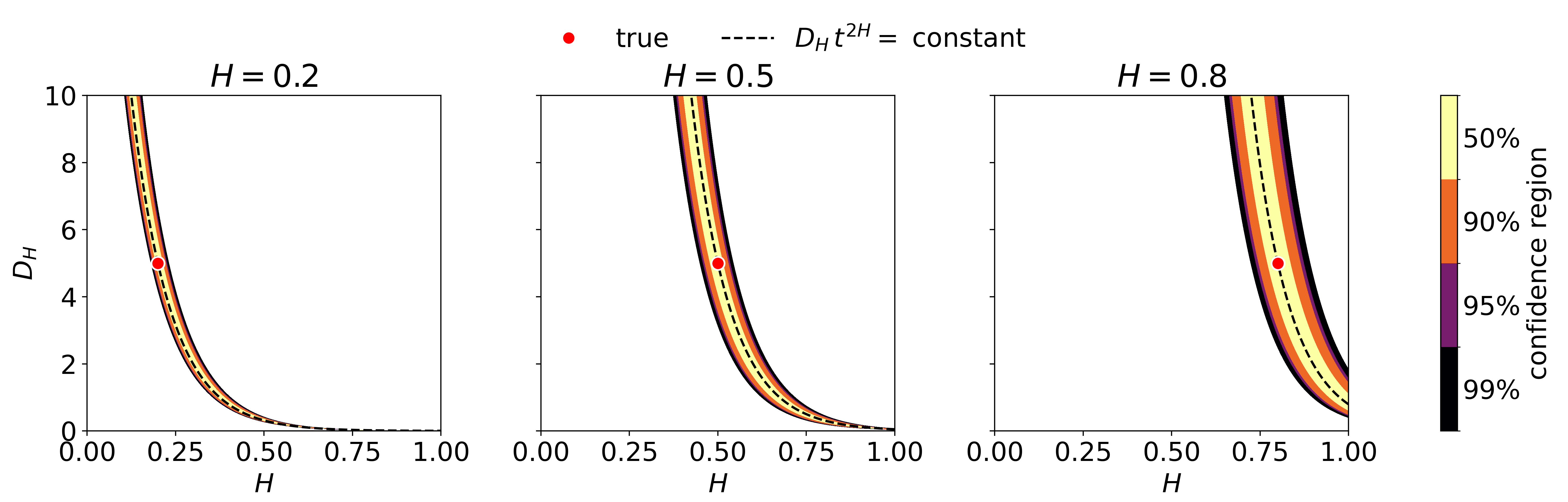}
    \caption{The loglikelihood landscapes for fBM PDFs with Poisson noise \eqref{eq:loglikelihood_poisson} against $D_H$ and $H$, given $N^o(x,t)$ at only a single time point $t=100$ with $\nu=500$, $S=1$ and $\lambda=0$, in bins of width $\Delta x=0.4$ over $|x|\leq20$. The landscape shown is for a single realisation of the Poisson noise model \eqref{eq:PoissonNoiseModel}, so $\ell_{\max}$ is specific to this realisation and may not lie at the true parameters. The colours show the loglikelihood relative to that maximum, $\ell-\ell_{\max}$. The coloured regions show $50$, $90$, $95$ and $99\%$ confidence regions for the parameters $(H,D_H)$, i.e. the sets $\{\theta: \ell_{\max}-\ell(\theta)\leq\tfrac{1}{2}\chi^2_2(p)\}$ \cite{wilks1938large}. The red dot shows the true parameters ($D_H = 5$) and the black dashed line shows the non-identifiability ridge \eqref{eq:identifiabilityEqPDF}.}
    \label{fig:loglikelihoodLandscape_pdfH_multiT100_nsamp1}
\end{figure}

\begin{figure}
    \centering
    \includegraphics[width=\linewidth]{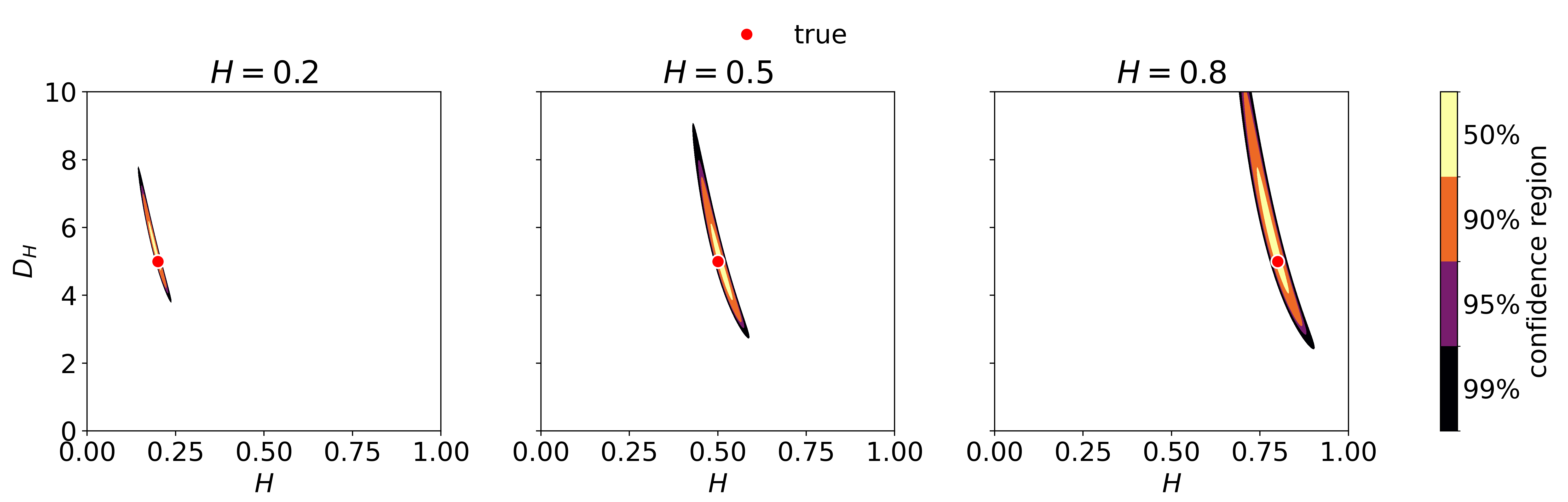}
    \caption{The loglikelihood landscapes for fBM PDFs with Poisson noise \eqref{eq:loglikelihood_poisson} against $D_H$ and $H$, given $N^o(x,t)$ at $K=10$ times $t_j = 10j$, $j=1,\dots,10$, with $\nu=500$, $S=1$ and $\lambda=0$, in bins of width $\Delta x=0.4$ over $|x|\leq20$. The confidence regions are the same as in Figure \ref{fig:loglikelihoodLandscape_pdfH_multiT100_nsamp1} and the red dot shows the true parameters ($D_H = 5$).}
    \label{fig:loglikelihoodLandscape_pdfH_multiT100_nsamp10}
\end{figure}

Figure \ref{fig:loglikelihoodLandscape_pdfH_multiT100_nsamp1} shows the loglikelihood landscapes \eqref{eq:loglikelihood_poisson} when the observed density is taken for a single time point ($t=100$).
If we denote the true parameter values as $D_H^*$ and $H^*$, then the parameters are non-identifiable as any pair of values $(H,D_H)$ that satisfies
\begin{equation}
    D_H t^{2H} = D_H^* t^{2H^*} = \text{constant},
    \label{eq:identifiabilityEqPDF}
\end{equation}
would maximise the loglikelihood given only one value of $t$.
This theoretical curve of equally likely parameter pairs is shown by the black dashed line in Figure~\ref{fig:loglikelihoodLandscape_pdfH_multiT100_nsamp1}.
However, if the observed density is available for multiple time points $t_1,t_2,\cdots,t_K$, then \eqref{eq:likelihood_poisson} and \eqref{eq:loglikelihood_poisson} become
\begin{equation}
    L(\theta,N|N^o) = \prod_{j=1}^{K} \prod_{i=1}^{M} \rho(\theta,N^o(x_i,t_j);N(x_i,t_j)),
    \label{eq:likelihood_poisson_nsamp}
\end{equation}
and
\begin{equation}
    \ell(\theta,N|N^o) = \sum_{j=1}^{K} \sum_{i=1}^{M} \ln(\rho(\theta,N^o(x_i,t_j);N(x_i,t_j))).
    \label{eq:loglikelihood_poisson_nsamp}
\end{equation}
Figure \ref{fig:loglikelihoodLandscape_pdfH_multiT100_nsamp10} shows the loglikelihood landscapes for $K=10$ with all other parameters being equal to Fig. \ref{fig:loglikelihoodLandscape_pdfH_multiT100_nsamp1}.
The progressively decreasing confidence regions in Fig. \ref{fig:loglikelihoodLandscape_pdfH_multiT100_nsamp10} closes around the true parameters such that $(H,D_H)$ becomes identifiable using multiple timepoints.
The regions are drawn from the likelihood ratio in the following way.
A parameter set $\theta$ is classified as within a $p$-level confidence region when $\ell_{\max}-\ell(\theta) \leq \chi^2_2(p)/2$, where $\ell_{\max}$ is the maximum loglikelihood over the grid of tested parameters and $\chi^2_2(p)$ is the $p$ quantile of the chi-square distribution with two degrees of freedom \cite{wilks1938large}.
With only one time observation, the likelihood confidence regions are a strip along the ridge \eqref{eq:identifiabilityEqPDF}, because $\ell$ is constant along it.
This is a direct consequence of \eqref{eq:identifiabilityEqPDF} if multiple time observations are available as opposed to a single time.

We can also explain this through the Fisher information.
In \eqref{eq:fBM_counts_exact}, $(H,D_H)$ only enter through the variance $2D_Ht_j^{2H}$ or equivalently through its logarithm
\begin{equation}
    u_j = \ln D_H + 2H\ln t_j.
    \label{eq:density_uj}
\end{equation}
The expected count in bin $i$ at time $t_j$ is $\mu_{ij} = SN(x_i,t_j)+\lambda$, a function of $u_j$.
Using the chain rule, its derivatives with respect to $\ln D_H$ and $H$ are
\begin{equation}
    \frac{\partial \mu_{ij}}{\partial \ln D_H} = \frac{\partial \mu_{ij}}{\partial u_j},
    \qquad
    \frac{\partial \mu_{ij}}{\partial H} = 2\ln t_j\,\frac{\partial \mu_{ij}}{\partial u_j}.
\end{equation}
For Poisson counts given $S$ and $\lambda$, the Fisher information is
\begin{equation}
    \mathcal{I}(\ln D_H, H) = \sum_{j=1}^{K} \sum_{i=1}^{M} \frac{1}{\mu_{ij}}\left(\frac{\partial \mu_{ij}}{\partial u_j}\right)^2 \begin{pmatrix} 1 & 2\ln t_j \\ 2\ln t_j & 4\ln^2 t_j\end{pmatrix}.
    \label{eq:density_fisher_rank}
\end{equation}
Clearly, each time point contributes a matrix of rank one to the Fisher information and so a single time observation can only determine the magnitude of $u_1$ and not the direction that $u_1$ is invariant.
This direction is the non-identifiability ridge \eqref{eq:identifiabilityEqPDF} in Figure~\ref{fig:loglikelihoodLandscape_pdfH_multiT100_nsamp1}.
This has built some intuition about fBM parameter inference using density data.
Next, we quantify how accurate the estimates of $H$ and $D_H$ are using data from density (count) measurements.

\subsection{Estimation accuracy and precision}
\label{sec:densityAccuracyPrecision}

We simulate the process described by \eqref{eq:PoissonNoiseModel} and maximise \eqref{eq:loglikelihood_poisson_nsamp} numerically, repeating the experiment $10^5$ times for each combination of $H \in \{0.2,0.5,0.8\}$, $K \in \{1,5,10,50,100\}$, $S \in [0.5,1.5]$ and $\lambda \in [0,1]$ (both in steps of $0.1$), a total of $1815$ configurations and $1.8\times 10^{8}$ estimates.
The gain and the background form an $11\times11$ grid, so every pair $(H,K)$ is measured on the $121$ combinations of $(S,\lambda)$.
Throughout, the truth is $D_H=5$ with $\nu=500$ counts per observation time distributed over $101$ bins of width $\Delta x = 2$, and the $K$ observation times are evenly spaced in $(0,100]$.
Each optimisation uses the Limited-memory Broyden-Fletcher-Goldfarb-Shanno with Bounds (L-BFGS-B) \cite{byrd1995limited,zhu1997algorithm} algorithm on $(\log_{10} D_H, H)$ from four random starts within the search box $D_H \in [10^{-3},10^{2}]$, $H \in [0.01,1]$, with the best being retained.
Table \ref{tab:densityMLE} reports the resulting sampling distributions at $S=1$, $\lambda=0$.
For a single time point, $K=1$, the estimates lie along the non-identifiability curve \eqref{eq:identifiabilityEqPDF} rather than about the true values, and where they fall on it reflects the optimiser and the search box rather than any property of the MLE or the data, so these estimates are not tabulated.

\begin{table}
    \centering
    \caption{Maximum-likelihood estimates of $H$ and $D_H$ from density data, using $10^5$ independent trials per row with $D_H=5$, $\nu=500$, $S=1$ and $\lambda=0$. Each observation is a histogram of $\nu$ counts in bins of width $\Delta x = 2$ over $|x|\leq 100$, at $K$ evenly spaced times in $(0,100]$. Point estimates are the mean for $\hat H$ and the median for $\hat D_H$. The brackets give the central $95\%$ interval of the sampling distribution over the $10^5$ trials, which describes the spread of the estimator. We show the mean for $\hat H$ since it is symmetric but the median for $\hat D_H$ since it is right-skewed as a positive scale parameter.}
    \label{tab:densityMLE}
    \begin{tabular}{ c | c | c | c  }
    $H$ & $K$ & $\hat{H}$ (mean) & $\hat{D}_H$ (median) \\
    \hline
    \hline
    \multirow{4}{*}{$0.2$} & 5 & $0.2003\;[0.152,\,0.250]$ & $4.983\;[3.36,\,7.34]$ \\
        & 10 & $0.2001\;[0.172,\,0.228]$ & $4.996\;[4.01,\,6.22]$ \\
        & 50 & $0.2001\;[0.190,\,0.210]$ & $4.998\;[4.63,\,5.39]$ \\
        & 100 & $0.2000\;[0.193,\,0.207]$ & $4.999\;[4.75,\,5.26]$ \\
    \hline
    \multirow{4}{*}{$0.5$} & 5 & $0.5003\;[0.451,\,0.550]$ & $4.993\;[3.35,\,7.36]$ \\
        & 10 & $0.5002\;[0.472,\,0.529]$ & $4.997\;[4.00,\,6.21]$ \\
        & 50 & $0.5000\;[0.490,\,0.510]$ & $4.999\;[4.63,\,5.39]$ \\
        & 100 & $0.5000\;[0.493,\,0.507]$ & $4.999\;[4.75,\,5.26]$ \\
    \hline
    \multirow{4}{*}{$0.8$} & 5 & $0.8006\;[0.726,\,0.877]$ & $4.975\;[2.87,\,8.62]$ \\
        & 10 & $0.8005\;[0.760,\,0.841]$ & $4.985\;[3.78,\,6.56]$ \\
        & 50 & $0.8001\;[0.787,\,0.814]$ & $4.998\;[4.58,\,5.45]$ \\
        & 100 & $0.8000\;[0.791,\,0.809]$ & $5.000\;[4.72,\,5.29]$ \\
    \hline
    \hline
    \end{tabular}
\end{table}

\begin{figure}
    \centering
    \includegraphics[width=\linewidth]{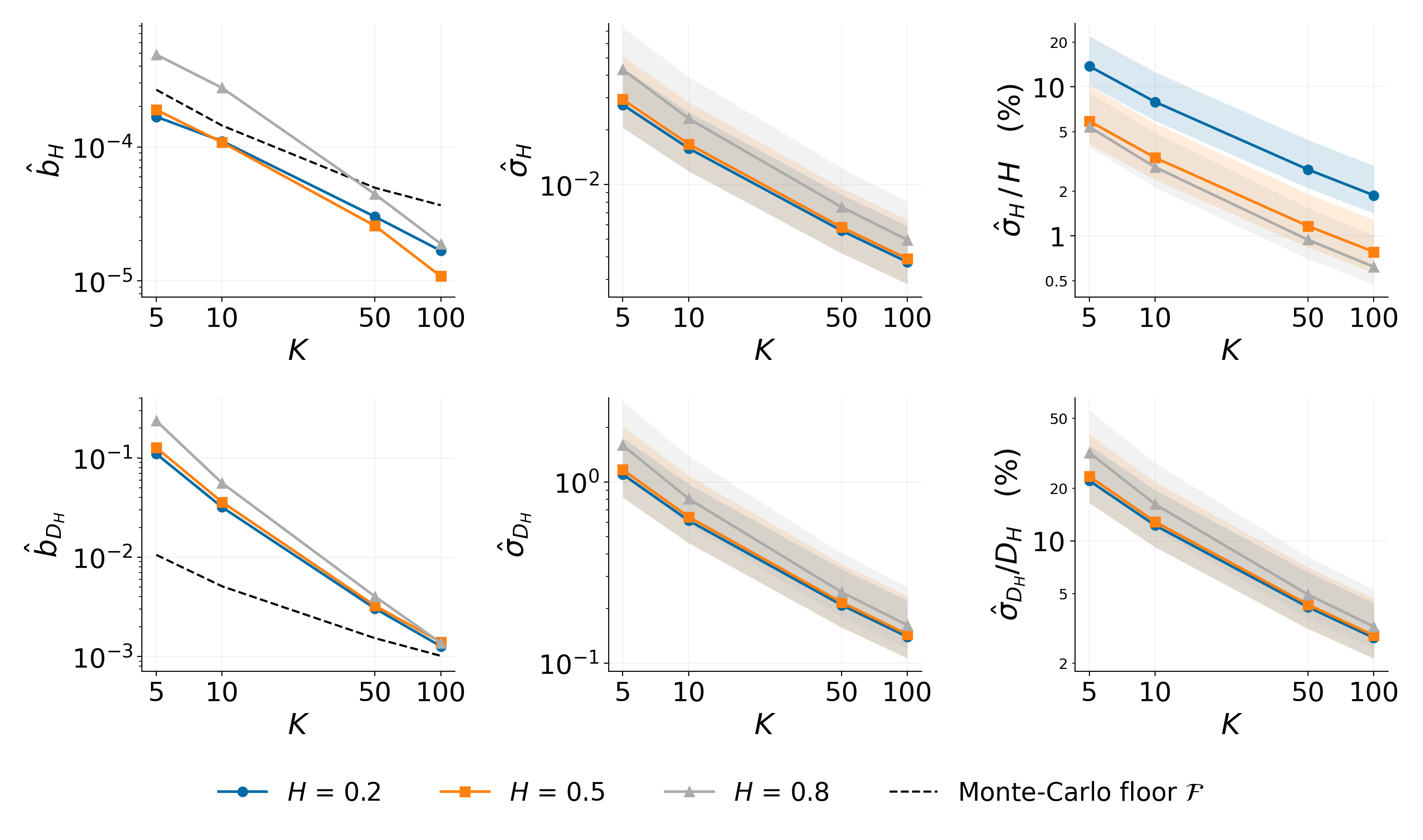}
    \caption{Accuracy and precision of the density MLE against the number of observation times $K$, for $\hat H$ (\textit{top row}) and $\hat D_H$ (\textit{bottom row}). All axes are logarithmic. The columns show the bias $\hat b_{\theta}$ \eqref{eq:mcBias} (\textit{left}), the nIQR $\hat\sigma_{\theta}$ \eqref{eq:mcIQR}, as a raw value (\textit{centre}) and as a percentage of the value being estimated (\textit{right}). Every quantity comes from $10^5$ repeated experiments per configuration. Each line is the median over the $121$ possible combinations of gain and background $(S,\lambda)$. The shaded bands (\textit{centre} and \textit{right}) show the maximum and minimum values for each median line. The black dashed lines (\textit{left}) show the bias detection floor $\mathcal{F}_{\theta}(K)$ \eqref{eq:mcFloor}.}
    \label{fig:optimisation_trials}
\end{figure}

\begin{figure}
    \centering
    \includegraphics[width=\linewidth]{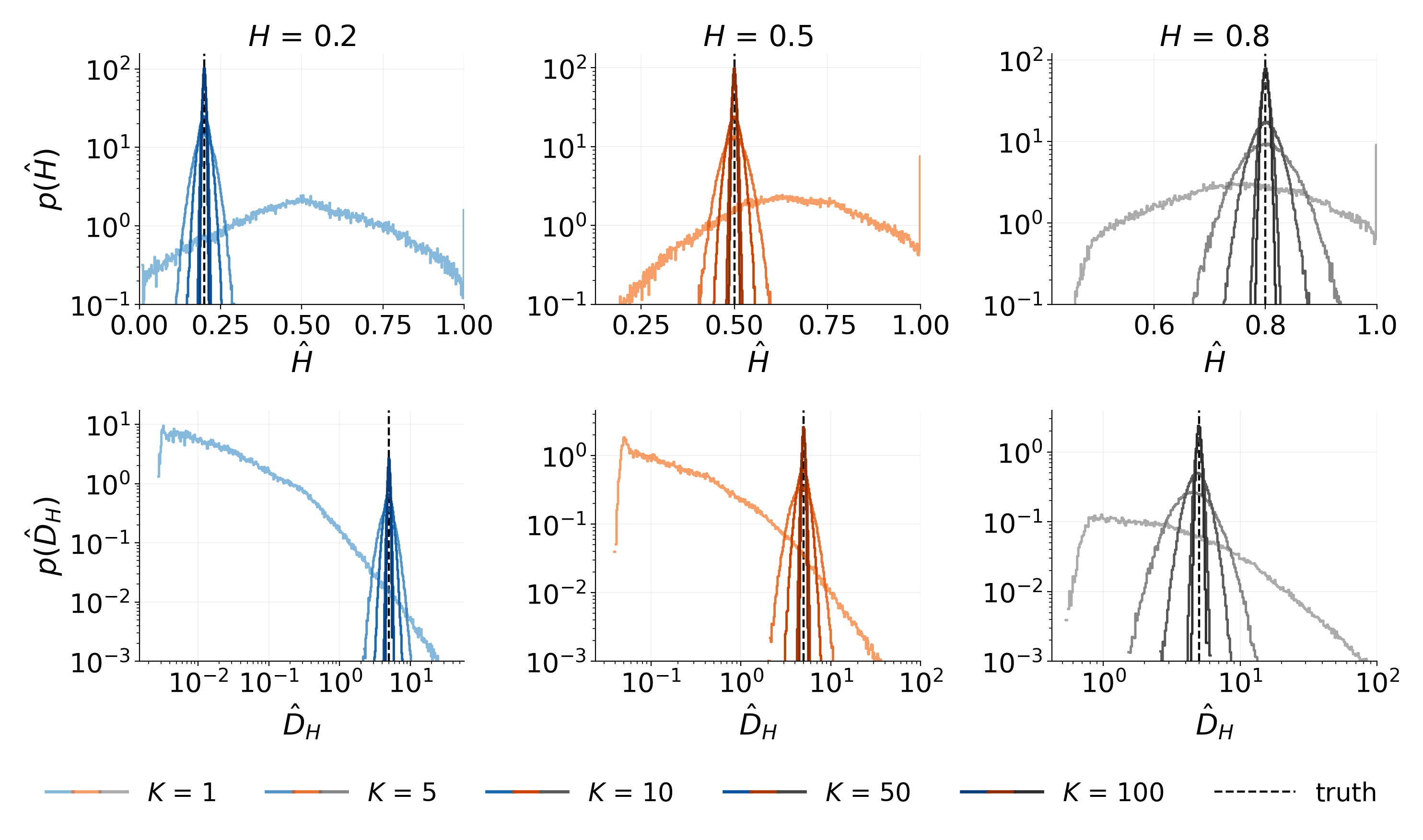}
    \caption{Sampling distributions of the density MLE at $S=1$, $\lambda=0.5$, for $H=0.2$, $0.5$ and $0.8$ (\textit{columns}). There are $n=10^5$ trials per curve. The rows show $\hat H$ (\textit{top}) and $\hat D_H$ (\textit{bottom}), with the truth marked with the black vertical dashed line. Shading marks the number of observation times $K$, from the lightest ($K=1$) to darkest ($K=100$).}
    \label{fig:optimisation_trials_distributions}
\end{figure}

Next, we quantify the accuracy and precision of parameter estimates using density data.
We write $\hat\theta_j$ for the estimate returned by the $j$th trial of $n = 10^5$ repeated experiments, where $\theta$ stands for either $H$ or $D_H$.
In the standard way, we measure accuracy by the bias
\begin{equation}
    \hat b_{\theta} \;=\; \frac{1}{n}\sum_{j=1}^{n}\hat\theta_j \;-\; \theta_{\mathrm{true}},
    \label{eq:mcBias}
\end{equation}
and we measure precision by the normalised interquartile range (nIQR) \cite{rousseeuw1993alternatives,hampel1974influence,analytical1989robust,analytical1989robust2}
\begin{equation}
    \hat\sigma_{\theta} \;=\; \frac{Q_{0.75}(\theta)-Q_{0.25}(\theta)}{1.349},
    \label{eq:mcIQR}
\end{equation}
where $Q_p(\theta)$ is the $p$th quantile of the $n$ estimates of $\theta$ and $Q_{0.75}(\theta)-Q_{0.25}(\theta)$ is the interquartile range.
The constant in the denominator of \eqref{eq:mcIQR} is the interquartile range of a standard normal distribution, $2\Phi^{-1}(3/4) = 1.34898$, in which $\Phi$ is the standard normal distribution function.
We use the nIQR because it is a robust measure for the spread of a random variable, which equals the standard deviation only when the sampling distribution is Gaussian.
Intuitively, the bias measures how centred the parameter estimates are on the true value and the standard deviation measures the spread of parameter estimates about the centre.
Finally, the sample standard deviation of the $n$ estimates is
\begin{equation}
    s_{\theta} \;=\; \sqrt{\frac{1}{n-1}\sum_{j=1}^{n}\big(\hat\theta_j - \langle \hat\theta \rangle \,\big)^{2}},
    \qquad \langle \hat\theta \rangle \;=\; \frac{1}{n}\sum_{j=1}^{n}\hat\theta_j.
    \label{eq:mcSD}
\end{equation}
The sample standard deviation determines if the estimation bias is real or an artefact of the finite number of repeat simulations.
Averaging $n$ estimates has an uncertainty of $s_{\theta}/\sqrt{n}$, so the measured bias is itself a random quantity.
Any measured bias smaller than $\pm 1.96\, s_{\theta}/\sqrt{n}$ is therefore not distinguishable from noise  $95\%$ of the time.
Since, our measurements involve four parameters, ($H, K, S, \lambda$), we fix the bias detection floor as
\begin{equation}
    \mathcal{F}_{\theta}(K) \;=\; 1.96 \max_{H} \; \operatorname*{med}_{(S,\lambda)} \; \frac{s_{\theta}}{\sqrt{n}},
    \label{eq:mcFloor}
\end{equation}
where the median value of the uncertainty is taken for the 121 measurement devices (combinations of $S$ and $\lambda$), and then the maximum of this median value for $H$.
This is a conservative floor and $\mathcal{F}_{\theta}(K)$ is shown by black dashed lines in the left panels of Figure \ref{fig:optimisation_trials}.
In both $H$ and $D_H$, Figure \ref{fig:optimisation_trials} shows that the bias in estimation becomes negligible as the number of time points sampled $K$ increases, which is consistent with the estimation method being derived from maximising likelihoods.
Figure \ref{fig:optimisation_trials} also quantifies the nIQR of estimation values $\hat{\sigma}_\theta$ as a raw value and as a percentage of the true value.
Figure \ref{fig:optimisation_trials_distributions} illustrates the distribution of estimates for values of $H$ and $D_H$ at fixed $(S,\lambda)$ for different $K$.

\subsection{Heuristics for gain, background noise and observation times}
\label{sec:density_heuristics}

\begin{figure}
    \centering
    \includegraphics[width=\linewidth]{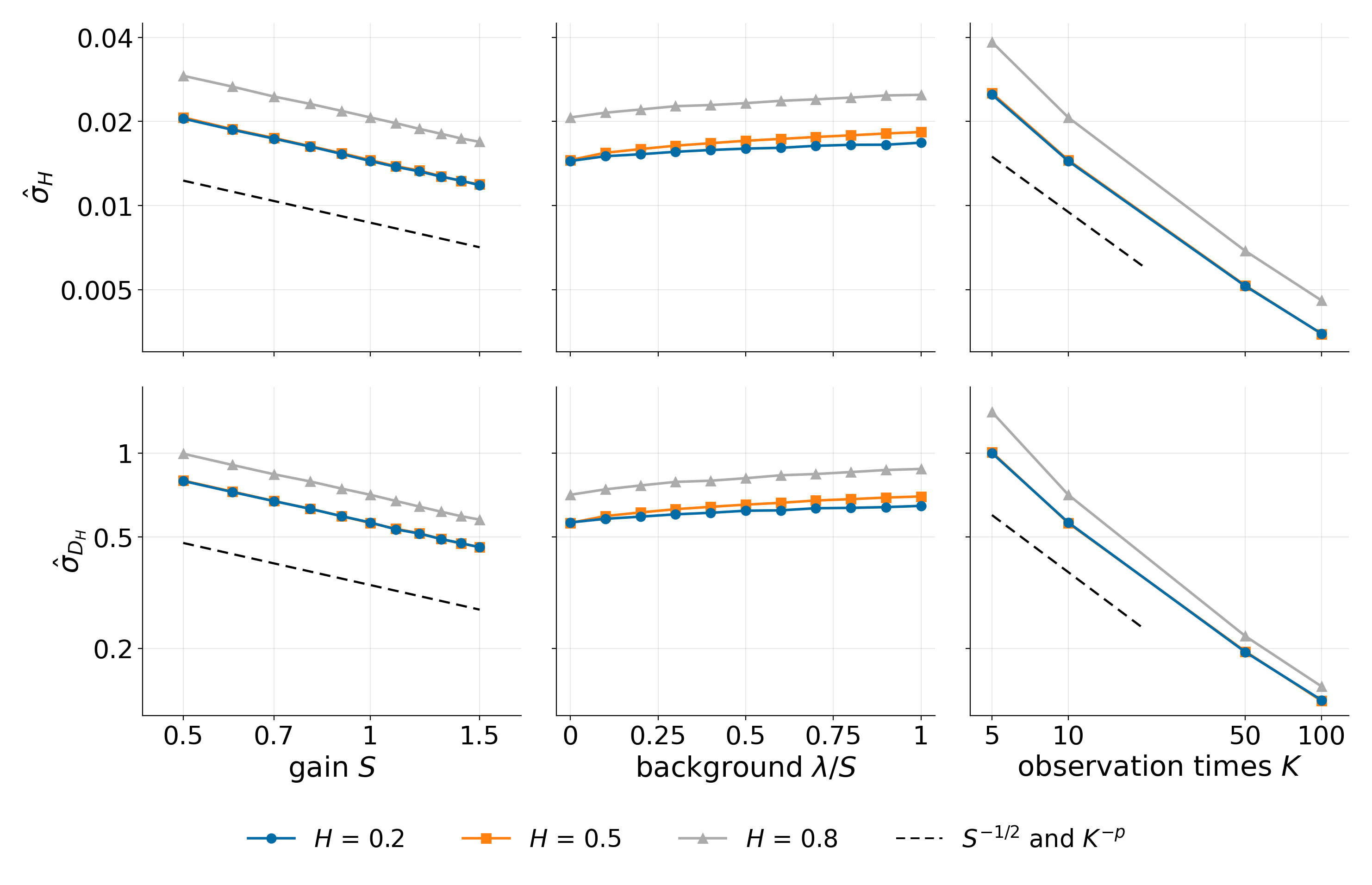}
    \caption{Gain $S$, background noise amplitude $\lambda$ and the number of observation times $K$ against the nIQR $\sigma_\theta$ \eqref{eq:mcIQR}, for $\hat H$ (\textit{top row}) and $\hat D_H$ (\textit{bottom row}). Each column varies one variable about the common reference device $K=10$, $S=1$, $\lambda=0$, holding the other two fixed. The plot uses $n=10^5$ repeated experiments per point. The black dashed lines show $\sigma_\theta \sim S^{-1/2}$  (\textit{left column}) and $\sigma_\theta \sim K^{-p}$, with $p=0.658$ (\textit{top right column}) and $p=0.678$ (\textit{bottom right column}).}
    \label{fig:optimisation_trials_collapse}
\end{figure}

For experiments, the gain $S$ and background noise $\lambda$ are determined by the measurement device while the number of observation times $K$ is set by experiment design.
Since the mean count of the $i$\textsuperscript{th} bin defined in \eqref{eq:PoissonNoiseModel} is $\mu_i = S N_i +\lambda$ and neither $S$ nor $\lambda$ depends on model parameters, the Fisher information, given $S$ and $\lambda$, can be written as
\begin{equation}
    \mathcal{I}_{ab}(\theta_{PDF};S,\lambda)=\sum_i \frac{(\partial_a\mu_i)(\partial_b\mu_i)}{\mu_i} = S\sum_i \frac{(\partial_a N_i)(\partial_b N_i)}{N_i+\lambda/S} \equiv S \mathcal{J}_{ab}(\theta_{PDF};\lambda/S),
    \label{eq:FisherInfoPoisson}
\end{equation}
where $a,b \in \{H,D_H\}$.
So the gain multiplies the entire information matrix while the amplitude of background noise enters only through the ratio $\lambda/S$.
Equation \eqref{eq:FisherInfoPoisson} is valid only when the gain and background are supplied (by calibration or otherwise) and every fit in this section assumes this.
In addition, the parameter $S$ absorbs the total number of particles $\nu$ if the total number of particles is also unknown.
Equation \eqref{eq:FisherInfoPoisson} implies that the standard error of the density MLE should decrease as $S^{-1/2}$ and this is shown on the left panels of Figure \ref{fig:optimisation_trials_collapse}.

Figure \ref{fig:optimisation_trials_collapse} varies $S$, $\lambda$ and $K$ about a common reference device and design ($K=10$, $S=1$, $\lambda=0$), plotting the resulting nIQR $\sigma_\theta$ \eqref{eq:mcIQR} of the estimates for $\hat H$ (\textit{top row}) and $\hat D_H$ (\textit{bottom row}).
Figure \ref{fig:optimisation_trials_collapse} verifies that the standard error of the density MLE decreases as $S^{-1/2}$ predicted by taking the square root of the diagonal inverse Fisher information matrix elements \eqref{eq:FisherInfoPoisson}.
It also shows that the nIQR of the density MLE degrades steadily as $\lambda/S$ increases from the zero-noise condition $\lambda=0$, but the effect is the smallest among the parameters compared.
Finally, the number of observation times $K$ has the strongest influence on MLE precision.
The right column of Figure \ref{fig:optimisation_trials_collapse} shows that $\hat\sigma_{\theta} \propto K^{-p}$  (approximately) with $p=0.658$ for $H$ and $p=0.678$ for $D_H$.
The fitted values are $p = 0.654$, $0.658$ and $0.705$ for $\hat H$ and $p = 0.676$, $0.678$ and $0.748$ for $\hat D_H$, at $H = 0.2$, $0.5$ and $0.8$, respectively.
Figure \ref{fig:optimisation_trials_collapse} heuristically shows, at least for the parameter ranges explored, that improvements should be prioritised first by number of observation times, then signal gain and finally background noise.
Finally, it is important to note that often, $K$ is an experimental design parameter that is both linked with the acquisition speed of the measurement device as well as being connected with the signal gain $S$.
For a camera, the maximum frame rate of observation is a parameter limited by the shutter speed, which also affects the gain $S$ needed to acquire an image.

Two limitations should be kept in mind when interpreting these results.
First, the domain size is held constant with $|x|\leq 100$ while the width of the density increases as $H$ increases.
At $t=100$ the domain therefore retains $\mathrm{erf}\big(x_{\max}/\sqrt{4D_Ht^{2H}}\big) = 100\%$, $99.8\%$ and $57\%$ of the $\nu$ counts at $H = 0.2$, $0.5$ and $0.8$.
This is equivalent to a fixed field of view of an acquisition device, which is an inherent property of experimental systems but leads to truncations of density measurements.
Second, the bin widths are fixed at $\Delta x = 2$ such that the number of resolved bins varies with $H$ and $t$.
While two factors do not introduce any bias, both are reasonable constraints for real experimental systems and it is the reason why different values of $H$ generate different amounts of data that ultimately can be used in estimation.

\section{MLE for Trajectories}
\label{sec:MLEtrajectories}

\begin{figure}
    \centering
    \includegraphics[width = \linewidth]{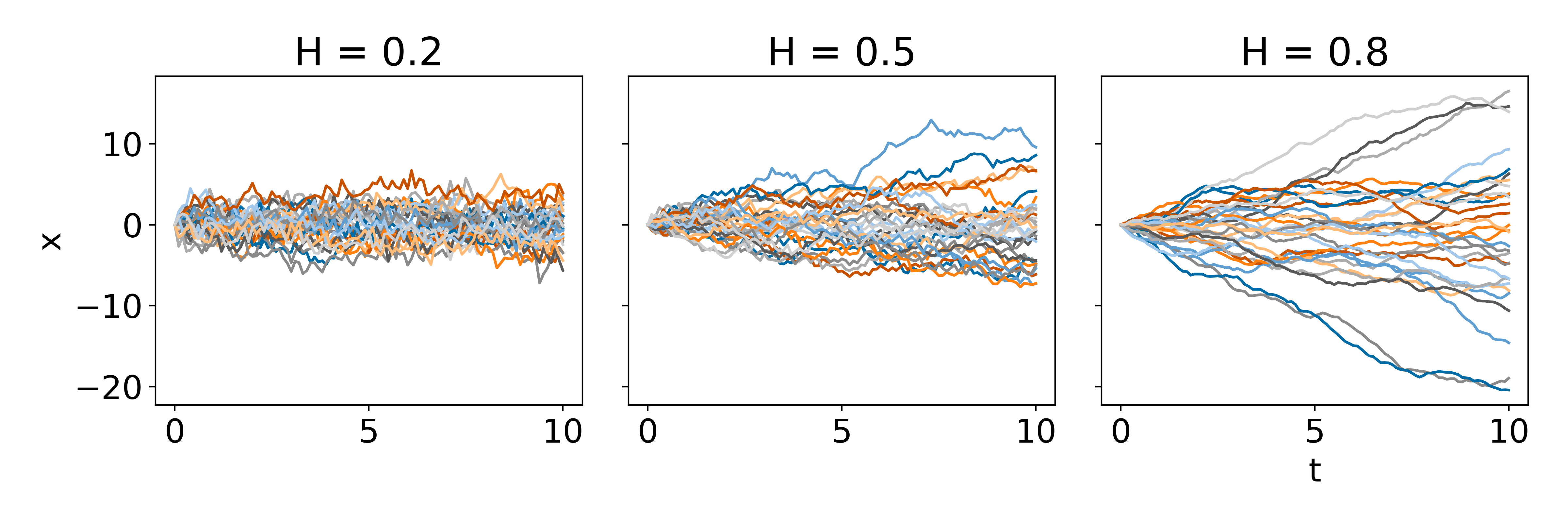}
    \caption{Trajectories for fBM with subdiffusive ($H=0.2$), diffusive ($H=0.5$) and superdiffusive ($H=0.8$) Hurst exponent up to time $t=10$ with $D_H = 1$}
    \label{fig:fbm_trajectories}
\end{figure}

In many cases, more detailed experimental data is available through tracked positions of discrete objects, such as individual intracellular vesicles or single proteins.
When dealing with data at this finer resolution, we require MLE using individual particle trajectories instead, which uses similar equations as the previous section.
However, instead of an observed particle count, $N^o(x_i,t)$, in the $i$\textsuperscript{th} position bin interval, we now have observed positions, $x_i^o(t)$, for particle $i$ at time $t$.
Figure \ref{fig:fbm_trajectories} shows trajectories of particles performing BM and fBM with initial condition $p(x,0)= \delta(x)$ for different Hurst exponents $H$.
The fBM model provides the expected density \eqref{eq:fBMDensity}, $p(x,t;\theta)$, given model parameters $\theta$ and for a single trajectory at a single instant in time.

\begin{figure}
    \centering
    \includegraphics[width=0.6\textwidth]{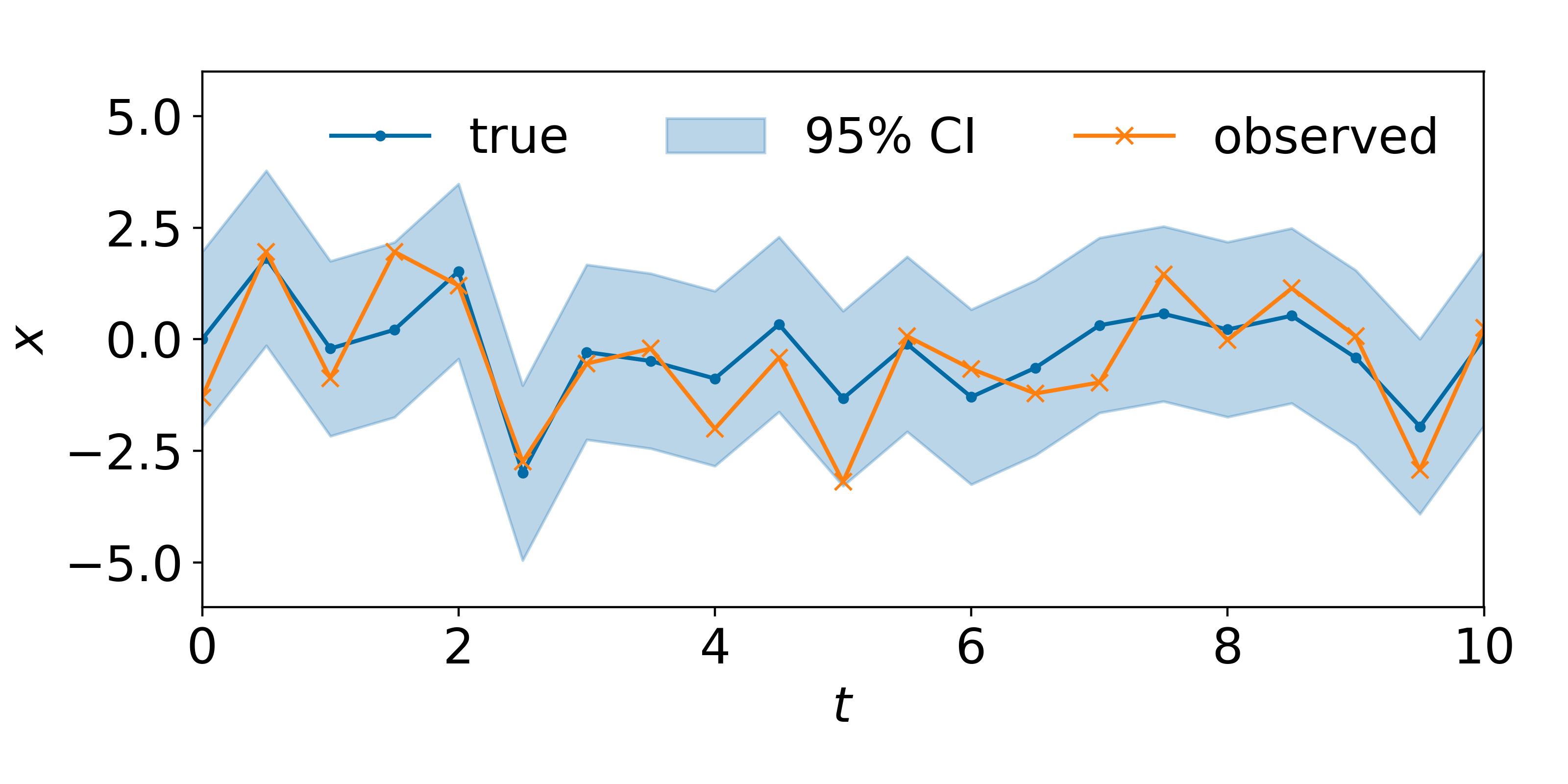}
    \caption{Plot of a single fBM trajectory, $x_i(t)$, (blue line) with $x_i(t)\pm1.96\sigma_{\epsilon}$ (blue shaded) and the observed trajectory, $x_i^o(t)$, (orange line) as given by \eqref{eq:fBM_noise} with $\sigma_{\epsilon}=1$, $D_H=1$ and $H=0.2$.}
    \label{fig:gaussianNoise_fBMtrajectory}
\end{figure}

Associated with each measurement of the true value $x_i(t_j)$, there is an additive Gaussian noise term such that the measured position is
\begin{equation}
    x_i^o(t_j) = x_i(t_j)+ \epsilon_{ij},
    \label{eq:fBM_noise}
\end{equation}
with $\epsilon_{ij} \sim \mathcal{N}(0,\sigma_{\epsilon}^2)$ independent across frames $j$ and particles $i$.
Such additive noise terms are typically introduced through localisation errors in tracking algorithms.
The PDF of the observed position, $x^o$, at a single instant in time is
\begin{equation}
    \begin{split}
        \rho(x^o,t) &= \int_{-\infty}^{\infty} p(x^o-\epsilon,t)\phi(\epsilon; 0,\sigma_{\epsilon}^2)d\epsilon\\
        &= \frac{1}{\sqrt{2\pi\left(2D_H t^{2H} + \sigma_{\epsilon}^2\right)}} \exp \left[-\frac{{x^o}^2}{2\left(2D_H t^{2H} + \sigma_{\epsilon}^2\right)}\right],
    \end{split}
    \label{eq:naive_traj_pdf}
\end{equation}
where
\begin{equation}
    \phi\left(\epsilon;\mu_{\epsilon},\sigma^2_{\epsilon}\right) = \frac{1}{\sqrt{2\pi \sigma^2_{\epsilon}}}\exp\left[-\frac{(\epsilon-\mu_{\epsilon})^2}{2\sigma_{\epsilon}^2}\right].
\end{equation}
Figure \ref{fig:gaussianNoise_fBMtrajectory} shows a realisation of a single true fBM trajectory and one that might be observed under experimental conditions with localisation error.

Unlike the likelihood for densities in the previous section, the likelihood for trajectories accounts for more information due to the covariance \eqref{eq:fbm_covariance}.
For a vector of positions $\mathbf{x}_i = \left(x_i(t_1), x_i(t_2),\cdots,x_i(t_K)\right)^T$ for the $i$\textsuperscript{th} particle over time, the PDF of the true trajectory is
\begin{equation}
    \rho(\mathbf{x}_i) = \frac{1}{\sqrt{(2\pi)^K \det \mathbf{\Sigma}}} \exp\left[  - \frac{1}{2} {\mathbf{x}_i}^T \mathbf{\Sigma}^{-1} \mathbf{x}_i  \right],
    \label{eq:exact_FBM_traj_pdf}
\end{equation}
where the covariance matrix is
\begin{equation}
    \left[\mathbf{\Sigma}(\theta)\right]_{ij} = D_H(t_i^{2H} + t_j^{2H} - |t_i-t_j|^{2H})
    \label{eq:fbm_covariance}
\end{equation}
for $1 \leq i,j\leq K$.
The PDF, $\rho(\mathbf{x}_i^o)$, of the observed trajectory \eqref{eq:fBM_noise} with additive noise is the same as \eqref{eq:exact_FBM_traj_pdf} except with covariance
\begin{equation}
    \left[\mathbf{\Sigma}(\theta)\right]_{ij} = D_H(t_i^{2H} + t_j^{2H} - |t_i-t_j|^{2H}) + \sigma_{\epsilon}^2 \delta_{ij},
    \label{eq:fbmNoise_covariance}
\end{equation}
where $\delta_{ij}$ is the Kronecker delta.

\begin{figure}
    \centering
    \includegraphics[width=\linewidth]{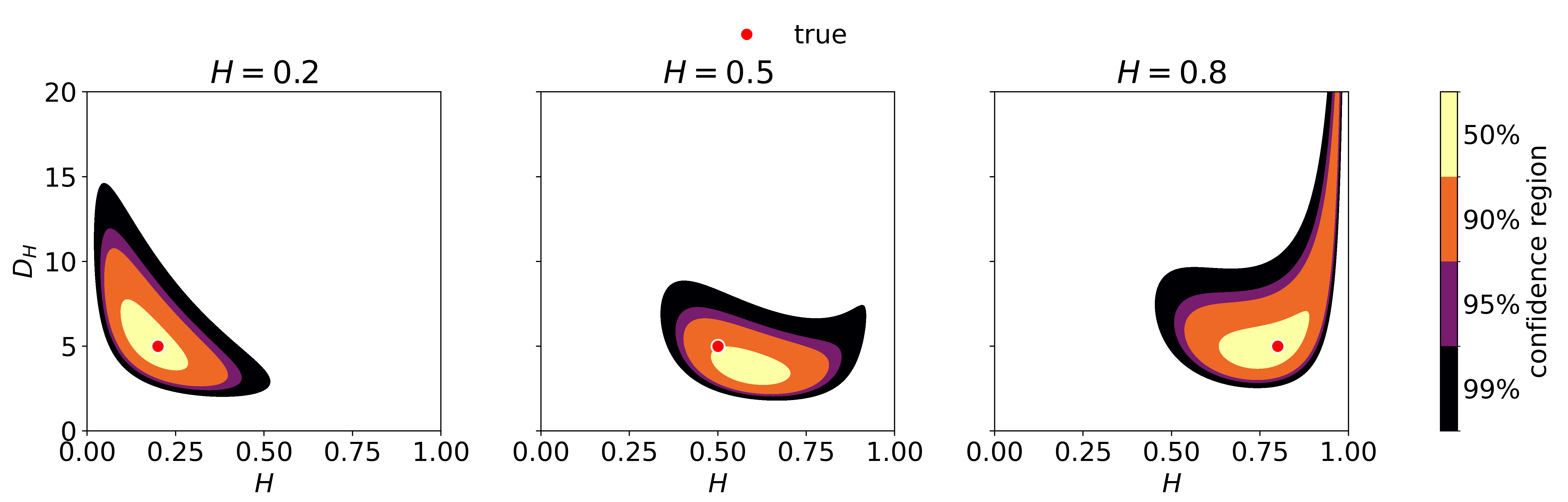}
    \caption{The loglikelihood landscapes for fBM trajectories with Gaussian noise \eqref{eq:fbm_traj_loglikelihood} against $D_H$ and $H$, for a single trajectory ($N_p=1$) with $K=50$, $D_H=5$, $t_K=100$ and $\sigma_{\epsilon}=1$ held at its true value in the fit. The coloured bands show $\ell-\ell_{\max}$, with $\ell_{\max}$ the maximum value for one trajectory. The different colours correspond to the $50$, $90$, $95$ and $99\%$ confidence regions as in Figure \ref{fig:loglikelihoodLandscape_pdfH_multiT100_nsamp1}. The red dot shows the true parameters.}
    \label{fig:loglikelihoodLandscape_trajH}
\end{figure}

The likelihood of observing $N_p$ trajectories with $K$ time points is
\begin{equation}
    L\left(\theta | \rho\right) = \prod_{i=1}^{N_p} \rho(\mathbf{x}_i^o),
    \label{eq:fbm_traj_likelihood}
\end{equation}
and the loglikelihood is
\begin{equation}
    \ell\left(\theta | \rho\right) = \ln \left(L(\theta | \rho)\right) = \sum_{i=1}^{N_p}  \ln\left( \rho(\mathbf{x}^o_i) \right).
    \label{eq:fbm_traj_loglikelihood}
\end{equation}
Here, $\theta = (\theta_{PDF},\sigma_{\epsilon})$ is the set of parameters to be determined by maximising the likelihood, where $\theta_{PDF}=(D_H,H)$ are the parameters of the model PDF in \eqref{eq:exact_FBM_traj_pdf}.
Figure \ref{fig:loglikelihoodLandscape_trajH} shows the loglikelihood landscapes \eqref{eq:fbm_traj_loglikelihood} for varying values of $H$.
The landscapes created using trajectory data clearly show that the model parameters $(D_H, H)$ are identifiable from a single trajectory with multiple time points.
At $H=0.5$ and $H=0.8$ the upper quantile bands show a ridge towards $H\to1$, along which the likelihood flattens.

\subsection{Estimation accuracy and precision}
\label{sec:traj_accuracy_precision}

We simulate the observed trajectory model \eqref{eq:fBM_noise} and maximise \eqref{eq:fbm_traj_loglikelihood} numerically to estimate the parameters $(H,D_H)$ for individual trajectories.
We perform $n=10^5$ optimisation trials for randomly generated trajectories for each combination of $H \in \{0.2,0.5,0.8\}$, $D_H \in \{0.5,5,50\}$, $\sigma_{\epsilon} \in \{0,0.5,1,3\}$ and $K \in \{5,10,50,100\}$, a total of $144$ configurations and $1.44\times 10^{7}$ trials.
The number of observation times $K$ are evenly spaced in $(0,100]$ and $t=0$ is excluded because $x(0)=0$ is an assumed initial condition.
Each optimisation uses L-BFGS-B on $(\log_{10} D_H,H)$ and retained the best likelihood from four random starts within the bounds $D_H \in [10^{-3},500]$ and $H \in [0.01,0.99]$.
We optimise $\log_{10}D_H$ since $D_H$ spans decades while $H$ spans less than one, so an optimisation algorithm using single step size gradient descent for both is dominated by noise in $D_H$.
Throughout this section $\sigma_{\epsilon}$ is held at its true value during the fit
so what is measured is the accuracy and precision of the estimator itself.
As in Sec. \ref{sec:densityAccuracyPrecision}, accuracy is measured by the bias $\hat b_{\theta}$ \eqref{eq:mcBias}, precision by the nIQR $\hat\sigma_{\theta}$ \eqref{eq:mcIQR}, and $\mathcal{F}_{\theta}$ \eqref{eq:mcFloor} is the Monte-Carlo floor below which a measured bias is indistinguishable.

\begin{table}
    \centering
    \caption{Maximum-likelihood estimates of $H$ and $D_H$ from single trajectories, from $100\,000$ independent trajectories per row with $D_H=5$ and $\sigma_{\epsilon}=0$. Each trajectory is observed at $K$ evenly spaced times in $(0,100]$ and fitted on its own by maximising \eqref{eq:fbm_traj_loglikelihood}, with $\sigma_{\epsilon}$ held at its true value. Point estimates are the mean for $\hat H$ and the median for $\hat D_H$, whose sampling distribution is strongly right-skewed at small $K$; brackets give the central $95\%$ interval of the sampling distribution.}
    \label{tab:trajMLE}
    \begin{tabular}{ c | c | c | c  }
    $H$ & $K$ & $\hat{H}$ (mean) $[2.5\%,\,97.5\%]$ & $\hat{D}_H$ (median) $[2.5\%,\,97.5\%]$ \\
    \hline
    \hline
    \multirow{4}{*}{$0.2$} & 5 & $0.2351\;[0.010,\,0.804]$ & $4.834\;[0.0494,\,34.5]$ \\
     & 10 & $0.2036\;[0.010,\,0.627]$ & $5.136\;[0.347,\,22.2]$ \\
     & 50 & $0.1952\;[0.054,\,0.348]$ & $4.975\;[2.77,\,8.37]$ \\
     & 100 & $0.1974\;[0.100,\,0.299]$ & $4.963\;[3.62,\,6.66]$ \\
    \hline
    \multirow{4}{*}{$0.5$} & 5 & $0.4221\;[0.010,\,0.907]$ & $5.845\;[0.230,\,155.]$ \\
     & 10 & $0.4495\;[0.010,\,0.834]$ & $5.335\;[0.779,\,57.4]$ \\
     & 50 & $0.4888\;[0.296,\,0.666]$ & $5.002\;[3.09,\,7.94]$ \\
     & 100 & $0.4938\;[0.362,\,0.617]$ & $4.974\;[3.71,\,6.49]$ \\
    \hline
    \multirow{4}{*}{$0.8$} & 5 & $0.6519\;[0.010,\,0.976]$ & $6.395\;[0.907,\,368.]$ \\
     & 10 & $0.7120\;[0.182,\,0.954]$ & $5.868\;[1.81,\,48.7]$ \\
     & 50 & $0.7804\;[0.587,\,0.923]$ & $5.010\;[3.12,\,9.00]$ \\
     & 100 & $0.7901\;[0.655,\,0.903]$ & $4.900\;[3.17,\,8.77]$ \\
    \hline
    \hline
    \end{tabular}
\end{table}

Table \ref{tab:trajMLE} reports the sampling distributions at $D_H=5$ and $\sigma_{\epsilon}=0$.
A single trajectory of $K$ points contains much less information that the densities used in Table \ref{tab:densityMLE}, which is why the confidence intervals are wider.
We choose to keep the different levels of available information between the density and trajectory data because this is typical of experimental conditions.
The density data is often acquired from experiments where fluorescent particles are assumed to share laws of movement and the bulk concentration is the quantity of interest.
The trajectory data is acquired when particles are suspected to move heterogeneously and thus parameter inference of each individual trajectory is of interest.
Inference from trajectory data is an inherently harder problem as the information available is lower.
Due to this, it becomes increasingly important to understand how accurate and precise MLE parameter inference is for fBM when attempting to characterise $H$ and $D_H$ for single particle trajectories.

\begin{figure}
    \includegraphics[width=\linewidth]{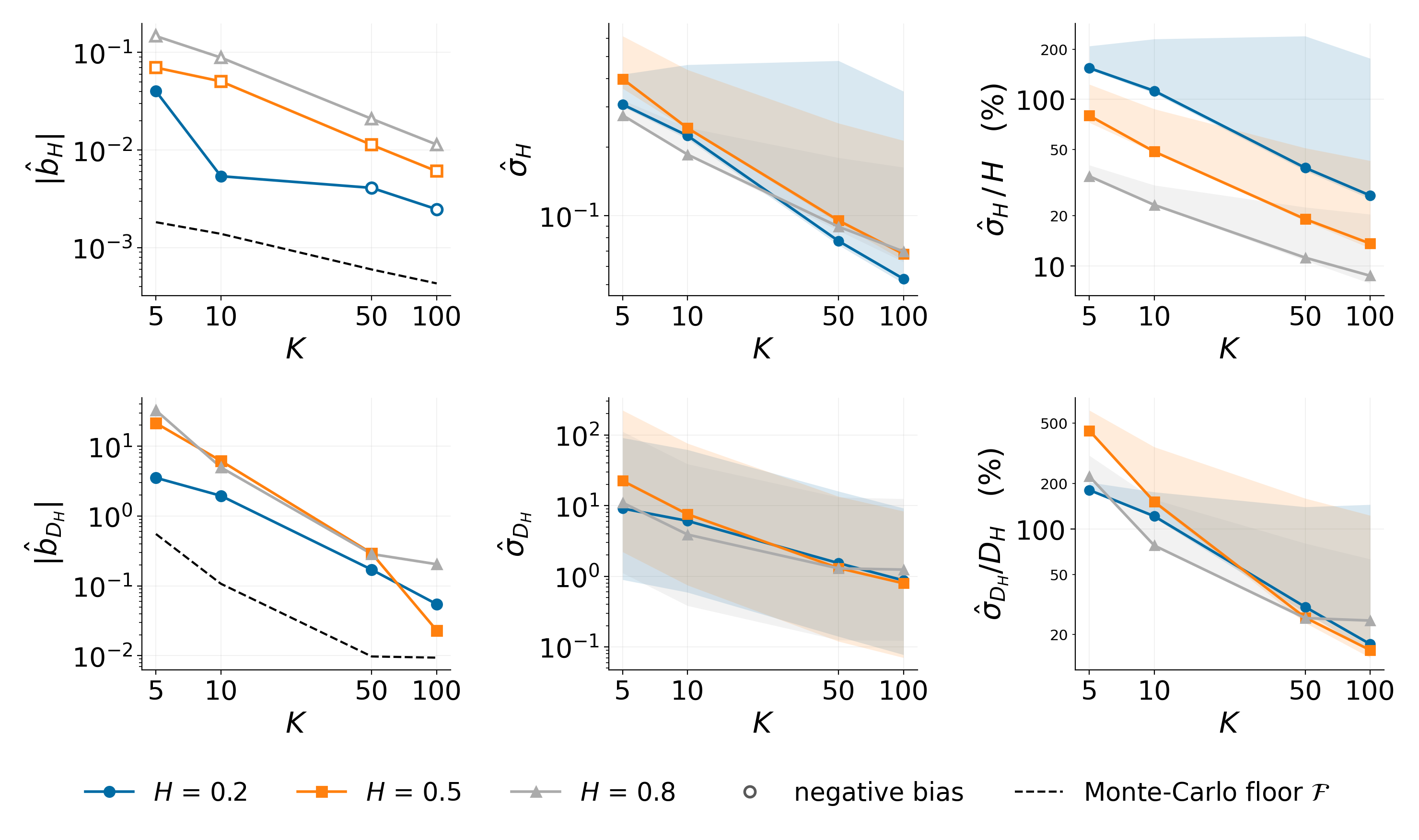}
    \caption{Accuracy and precision of the single-trajectory MLE against the number of observation times $K$, for $\hat H$ (\textit{top row}) and $\hat D_H$ (\textit{bottom row}).  The columns show the magnitude of the bias $|\hat b_{\theta}|$ \eqref{eq:mcBias} (\textit{left}), the nIQR $\hat\sigma_{\theta}$ \eqref{eq:mcIQR} as a raw value (\textit{centre}) and as a percentage of the value being estimated (\textit{right}). Every quantity is computed from $10^5$ estimates per configuration, each the maximum-likelihood fit to one simulated trajectory of $K$ points. Each line is the median over the $12$ combinations of the generalised diffusion coefficient and localisation error $(D_H,\sigma_{\epsilon})$; open markers denote a negative bias. The shaded bands (\textit{centre} and \textit{right}) show the maximum and minimum over those combinations. The black dashed lines (\textit{left}) show the bias detection floor $\mathcal{F}_{\theta}(K)$ \eqref{eq:mcFloor}.}
    \label{fig:traj_optimisation_trials}
\end{figure}

Similar to \ref{sec:densityAccuracyPrecision}, Figure \ref{fig:traj_optimisation_trials} shows the bias and nIQR against the number of time points $K$.
Unlike the density MLE of Sec. \ref{sec:MLEdensities}, there is a real bias in estimation for $H$ and $D_H$ using trajectory independent of the above the sample size noise shown in Figure \ref{fig:traj_optimisation_trials} (left).
Precision improves monotonically with $K$ shown in Figure \ref{fig:traj_optimisation_trials} (centre and right) and the relative precision $\hat\sigma_H/H$ (right) shows that estimation when $H = 0.2$ is a harder task.
This is despite the raw value of $\sigma_H$ being the lowest for $H=0.2$, similar to the density MLE in Figure \ref{fig:optimisation_trials} (right).
While the trajectory MLE may seem inferior to the density MLE, the comparison between the trajectory MLE and density MLE is not of equal information.
The density data sampled at $K$ spatial locations at multiple time points already has far greater information than a single trajectory sampled at $K$ time points.
However, the trajectory data instead contains the correlation between observations over time, which allows the trajectory MLE to identify $(H,D_H)$ from a single trajectory.

\begin{figure}
\includegraphics[width=\linewidth]{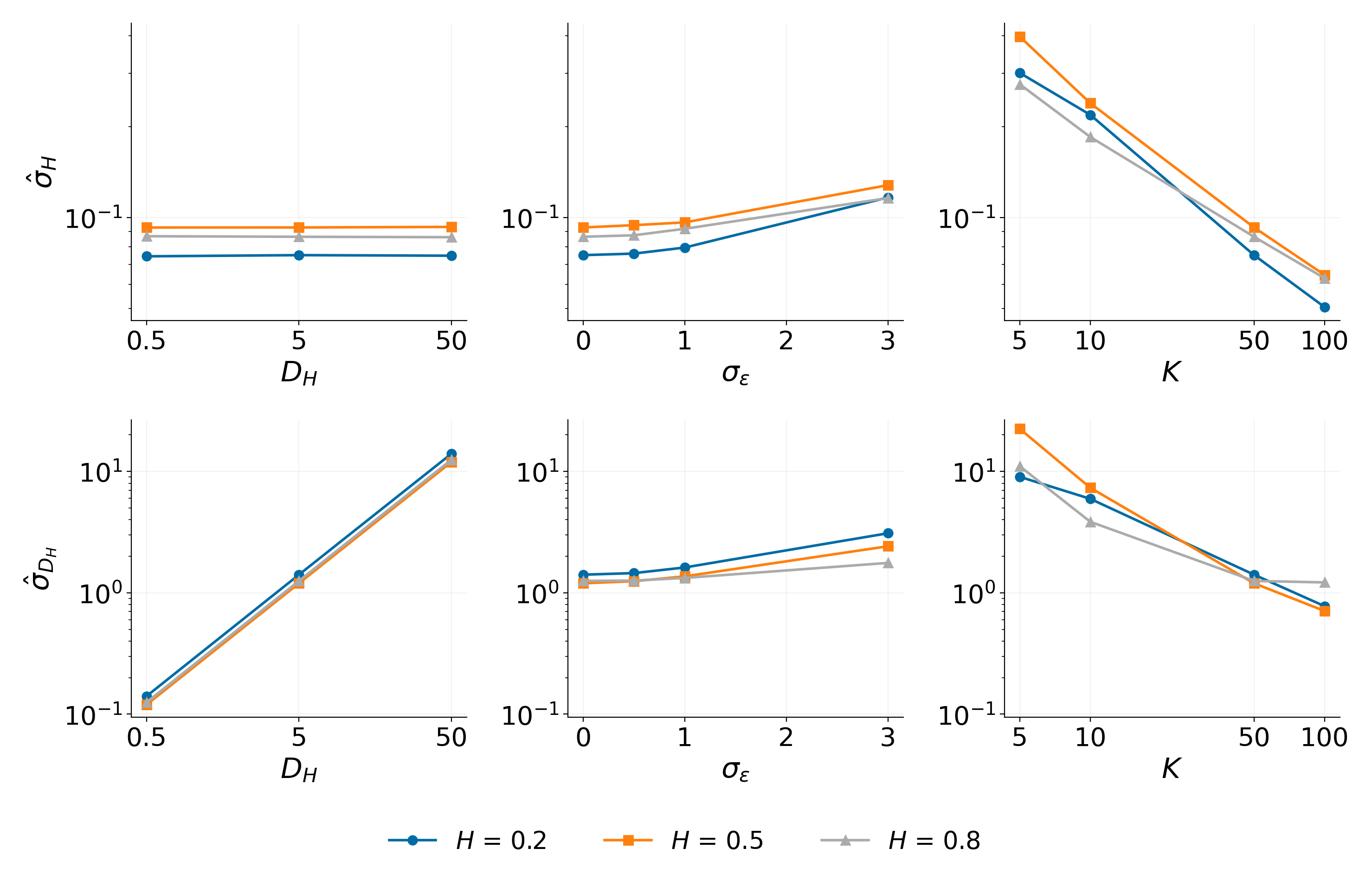}
\caption{Diffusion coefficient $D_H$, localisation error scale $\sigma_{\epsilon}$ and the number of observation times $K$ against the nIQR $\hat\sigma_{\theta}$ \eqref{eq:mcIQR}, for $\hat H$ (\textit{top row}) and $\hat D_H$ (\textit{bottom row}). Each column varies one variable about $D_H=5$, $\sigma_{\epsilon}=0$, $K=50$, holding the other two fixed.}
\label{fig:traj_knobs}
\end{figure}

Figure \ref{fig:traj_knobs} shows how the nIQR $\sigma_{\theta}$ varies as the diffusion coefficient $D_H$, localisation error scale $\sigma_{\epsilon}$ and the number of observation times $K$ changes.
Each plot shows this change as one parameter is varied about a common reference point ($D_H=5$, $\sigma_{\epsilon}=0$, $K=50$).
It is clear that the number of observation points $K$ is the most important factor for $\sigma_H$.
This is true also for $\sigma_{D_H}$ but the precision of estimating $D_H$ is also dependent on the magnitude of the value.
This magnitude dependence on $D_H$ disappears when considering the relative precision $\sigma_{D_H}/D_H$, when $\sigma_\epsilon=0$.
Localisation error degrades the precision of estimating both quantities but relatively less than the improvement gained by increasing $K$.
Note that when $\sigma_\epsilon=0$, $\sigma_H$ does not depend on $D_H$ at all, which can be derived from the Fisher information matrix or seeing that a rescaled position does not influence the loglikelihood \eqref{eq:fbm_traj_loglikelihood} apart from an additive constant.
With localisation error, the same rescaling shows that the distribution of $\hat H$ depends on $D_H$ and $\sigma_\epsilon$ only through the ratio $\sigma_\epsilon^2/D_H$.
The parameters $D_H$ and $\sigma_{\epsilon}$ and $K$ are usually difficult to change in an experimental setting that measures trajectories and presents a very different set of challenges compared to the density MLE case.
Of these, only $K$ is somewhat under the experimenter's control, through the frame rate of the acquisition device, and it also has the largest effect.
Even then, the number of time points $K$ in any given trajectory is influenced by the quality of the single particle tracking algorithm.
$D_H$ is a property of the system and usually used as an independent variable.
$\sigma_{\epsilon}$ is a property of the microscope and the tracker together, and it should be calibrated once, from immobilised particles or from the tracker's own residuals, and supplied to the fit rather than estimated from each trajectory.

\begin{figure}
\includegraphics[width=\linewidth]{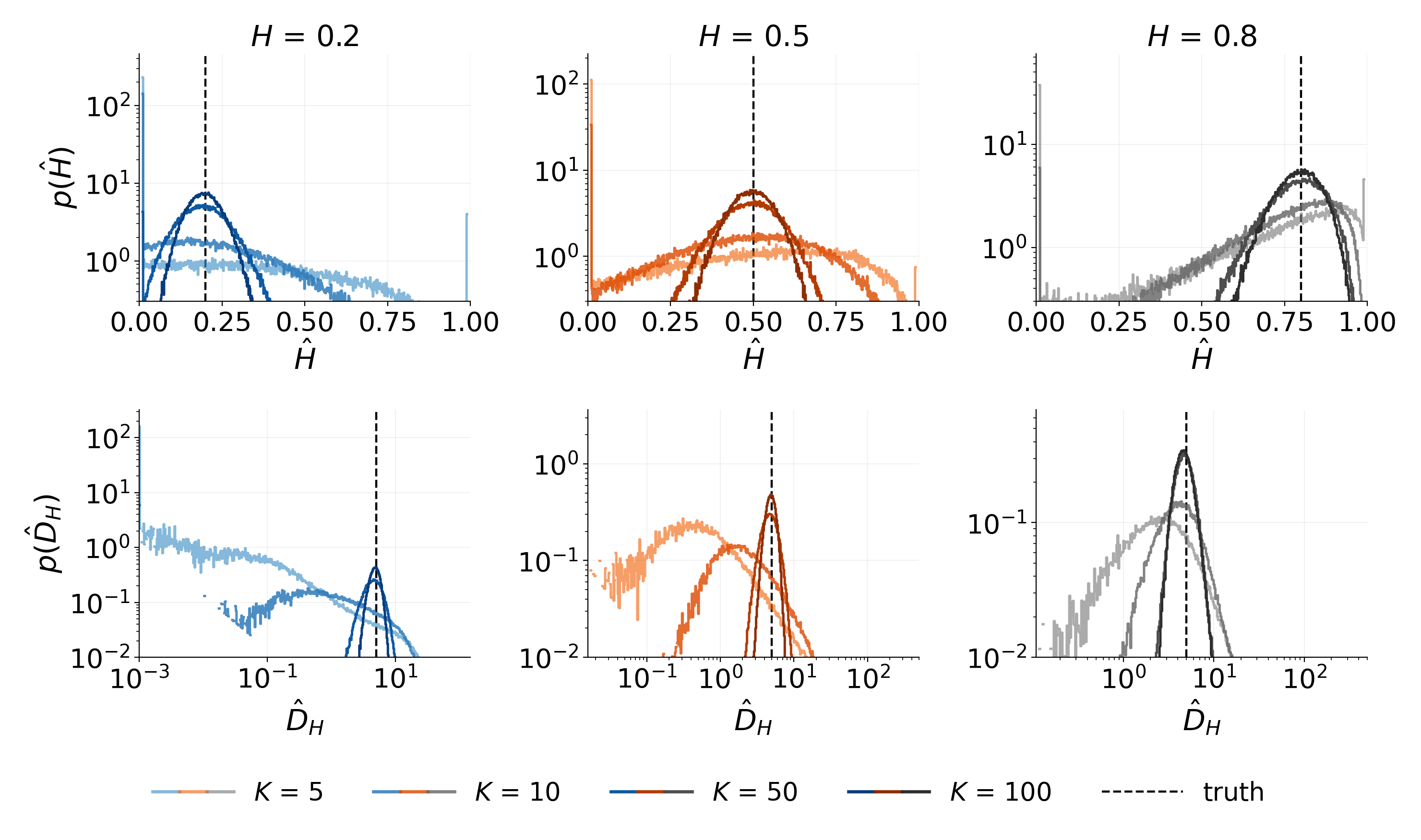}
\caption{Sampling distributions of the single-trajectory MLE at $D_H=5$ and $\sigma_{\epsilon}=1$, for $H=0.2$, $0.5$ and $0.8$ (\textit{columns}). Each curve is the distribution of $10^5$ estimates, each the maximum-likelihood fit to one simulated trajectory of $K$ points. The rows show $\hat H$ (\textit{top}) and $\hat D_H$ (\textit{bottom}), with the truth marked by the black vertical dashed line. Shading marks the number of observation times $K$, from the lightest ($K=5$) to the darkest ($K=100$).
When the sampling distribution exceeds the bounds of the numerical optimisation algorithm, the mass that falls outside the bounds are set as values at the bound.
}
\label{fig:traj_distributions}
\end{figure}

Figure \ref{fig:traj_distributions} shows the sampling distributions of the parameter estimates and demonstrates that the large nIQR $\sigma_{\theta}$ at small $K$ is not arising from a wide unimodal distribution.
Note for small $K$, a fraction of estimates sit at the bounds of $H$ at either 0.01 or 0.99.
The railing to the upper and lower bounds is explained by the covariance \eqref{eq:fbmNoise_covariance}.
From \eqref{eq:fbmNoise_covariance}, the dependence of $\mathbf{\Sigma}$ on $H$ is separable from $D_H$, which is a multiplicative scale factor, so we write
\begin{equation}
    \mathbf{\Sigma}(\theta) = D_H\,\mathbf{\Sigma}_0(H) + \sigma_{\epsilon}^2\mathbf{I},
    \qquad
    \left[\mathbf{\Sigma}_0(H)\right]_{ij} = t_i^{2H} + t_j^{2H} - |t_i-t_j|^{2H},
    \label{eq:unit_covariance}
\end{equation}
where $\mathbf{I}$ is the identity matrix.
When $\sigma_{\epsilon}=0$, the covariance is $\mathbf{\Sigma}=D_H\mathbf{\Sigma}_0(H)$, which gives $\det\mathbf{\Sigma}=D_H^K\det\mathbf{\Sigma}_0(H)$ and $\mathbf{\Sigma}^{-1}=\mathbf{\Sigma}_0(H)^{-1}/D_H$.
Then, the loglikelihood \eqref{eq:fbm_traj_loglikelihood} for a single trajectory $\mathbf{x}$ is $-\tfrac{K}{2}\ln 2 \pi-\tfrac{K}{2}\ln D_H-\tfrac{1}{2}\ln\det\mathbf{\Sigma}_0(H)-\mathbf{x}^T\mathbf{\Sigma}_0(H)^{-1}\mathbf{x}/(2D_H)$.
Maximising the loglikelihood over $D_H$ given $H$, we can write the closed form estimate for the generalised diffusion coefficient as
\begin{equation}
    \hat{D}_H(H) = \frac{1}{K}\,\mathbf{x}^T\mathbf{\Sigma}_0(H)^{-1}\mathbf{x}.
    \label{eq:profiled_amplitude_traj}
\end{equation}
Substituting \eqref{eq:profiled_amplitude_traj} into \eqref{eq:fbm_traj_loglikelihood} gives the profile loglikelihood in $H$,
\begin{equation}
    \begin{split}
    \ell_p(H) & = -\frac{K}{2}\ln\hat{D}_H(H) - \frac{1}{2}\ln\det\mathbf{\Sigma}_0(H) - \frac{K}{2} - \frac{K}{2}\ln 2\pi
\end{split}
    \label{eq:profile_loglikelihood}
\end{equation}
At $H=1$, $t_i^{2H} + t_j^{2H} - |t_i-t_j|^{2H} = 2t_it_j$, so
\begin{equation}
    \mathbf{\Sigma}_0(1) = 2\,\mathbf{t}\mathbf{t}^T,
    \label{eq:rank_one}
\end{equation}
which has rank one.
A ballistic trajectory has $x(t)=vt$ with constant $v$ such that $K$ observations over time reduce to one degree of freedom.
As $H\to1$, the $K-1$ eigenvalues of $\mathbf{\Sigma}_0$ vanish proportional to $1-H$ \cite{kato1966perturbation} and the middle term in \eqref{eq:profile_loglikelihood} increases as $-\tfrac{1}{2}(K-1)\ln(1-H)$.
Using the same argument, the quadratic form $\mathbf{x}^T\mathbf{\Sigma}_0(H)^{-1}\mathbf{x}$ grows like $1/(1-H)$ whenever $x(t) \neq vt$ and so as $H\to 1$, the first term in \eqref{eq:profile_loglikelihood} decreases as $\tfrac{1}{2}K\ln(1-H)$.
Overall, $\ell_p(H)\simeq\tfrac{1}{2}\ln(1-H)\to-\infty$ as $H\to1$.
Because $\ell_p(H)$ decays logarithmically, the difference in loglikelihood values at $H=0.9$ and $H=0.99$ is small enough that some optimisation trials end at the upper bound.
The same argument holds when $\sigma_{\epsilon}>0$, although \eqref{eq:profiled_amplitude_traj} no longer applies and $D_H$ must be maximised numerically.

On the other hand, $t^{2H}\to1$ as $H\to 0$ for every $t>0$ and $0^{2H} = 0$ for every $H>0$. So
\begin{equation}
    \lim_{H\to0^+}\mathbf{\Sigma}(\theta) = D_H\left(\mathbf{I} + \mathbf{1}\mathbf{1}^T\right) + \sigma_{\epsilon}^2\mathbf{I},
    \label{eq:covH0}
\end{equation}
where $\mathbf{1}$ is the vector of ones.
The only problem becomes an identifiability issue when both $D_H$ and $\sigma_{\epsilon}$ are being estimated together.
In Figure \ref{fig:traj_distributions}, the noise parameter $\sigma_{\epsilon}$ is given not estimated. Figure \ref{fig:traj_distributions} shows a significant number of estimates for $H$ at the lower bound for small $K$ and this is a truncation error from the bounds on the optimisation algorithm.

Here, we have documented both pathologies of the trajectory MLE and do not provide traditional statistical solutions for improving estimation, such as regularisation terms that may introduce a bias.
We also do not supply a method to remedy the non-identifiability of $D_H$ and $\sigma_{\epsilon}$ as the localisation error $\sigma_{\epsilon}$ can be calibrated in an experiment by imaging and tracking stationary particles.
In what follows, we introduce a hybrid estimation method that uses the neural network estimator for the Hurst exponent \cite{han2020deciphering} and a profile likelihood MLE for $D_H$.

\subsection{A hybrid likelihood and neural network estimation method}
\label{sec:estimator_comparison}

Many estimators for fBM focus on estimating the Hurst exponent $H$ while leaving $D_H$ as a scale factor to be determined afterwards.
We argue that estimating $H$ and $D_H$ together is critical since the likelihood for fBM is joint between the two.
Here, we provide a comparative analysis across different methods and find a hybrid likelihood and posterior mean estimation method that is better than MLE, except near $H=0$ and $H=1$.
We also include the exact posterior mean of $H$ under the prior the network was trained on.
Section~\ref{sec:posterior_mean} shows that this is the estimator the neural network approximates.
To estimate $H$, the methods we compare are the time-averaged mean-squared displacement (TAMSD) \cite{qian1991single,metzler2014anomalous,kepten2015guidelines}, the wavelet-variance estimator \cite{percival1995estimation,percival2000wavelet}, detrended fluctuation analysis (DFA) \cite{peng1994mosaic,kantelhardt2001detecting}, rescaled-range analysis (R/S) \cite{hurst1951long,mandelbrot1969robustness}, fractal dimension by Higuchi \cite{higuchi1988approach}, the MLE detailed in Section \ref{sec:traj_accuracy_precision} above and the neural network estimator variant of Han et al. \cite{han2020deciphering}.
To estimate $D_H$, we compare the TAMSD, MLE, the posterior mean with the closed-form amplitude \eqref{eq:profiled_amplitude_traj} (PM$+$pMLE) and a hybrid method (NN+pMLE).
The hybrid method, NN+pMLE, involves estimating $H$ using a neural network and then using the profile likelihood \eqref{eq:profiled_amplitude_traj} to evaluate $\hat{D}_H$.
This hybrid method with two stage estimation of $H$ and $D_H$ has been explored previously by using a recurrent network to estimate $H$ and a second recurrent network that estimates $D_H$ given $H$ \cite{korabel2025deep}.
The method we propose differs from this approach as we use the profile likelihood \eqref{eq:profiled_amplitude_traj} directly given $\hat{H}$.
This requires no secondary network and the estimate RMSE sits near the Cram\'er-Rao bound.
To compare the different methods, we use the same set of simulated fBM trajectories so that any differences are due to the estimator and not spurious differences in the pseudo-random numbers.
We exclude any localisation error, $\sigma_{\epsilon}=0$, because it should be calibrated for each experimental system used to generate data and supplied to the parameter estimation method directly, not inferred.
Practical implementation details are given in Section \ref{sec:estimator_implementation}.
All simulations below use $K=50$, $t_K=100$ and $D_H=5$ with statistics calculated from $10^5$ estimations on individual trajectories at each $H=\{0.2,0.5,0.8\}$.

\begin{table}
    \centering
    \begin{tabular}{l|ccc}
        Estimator & $H{=}0.2$ & $H=0.5$ & $H=0.8$ \\
        \hline
        MLE & 0.075 & 0.094 & 0.090 \\
        PM+pMLE & \textbf{0.074} & \textbf{0.091} & \textbf{0.081} \\
        NN+pMLE & 0.082 & 0.100 & 0.090 \\
        TAMSD & 0.087 & 0.121 & 0.135 \\
        Wavelet & 0.136 & 0.142 & {0.143} \\
        DFA & {0.188} & {0.169} & {0.179} \\
        R/S & {0.233} & {0.175} & {0.152} \\
        Higuchi & 0.094 & 0.110 & 0.113 \\
        \hline
        $\mathcal{B}_{H}$ & 0.073 & 0.090 & 0.083 \\
    \end{tabular}
    \caption{
        RMSE of $\hat H$ for estimators on one set of $10^5$ simulated trajectories without localisation error. The best (lowest RMSE) entry for each $H$ value is bold (see Section \ref{sec:estimator_implementation} for details).
        NN+pMLE is the exponent returned by the neural network alone and PM+pMLE is $\mathbb{E}[H\,|\,\mathbf{x}]$ under the network's training prior.
        The last row is the Cram\'er--Rao bound $\mathcal{B}_{H}$ from \eqref{eq:crb_H}.
    }
    \label{tab:estimator_comparison}
\end{table}

\begin{figure}
    \includegraphics[width=\linewidth]{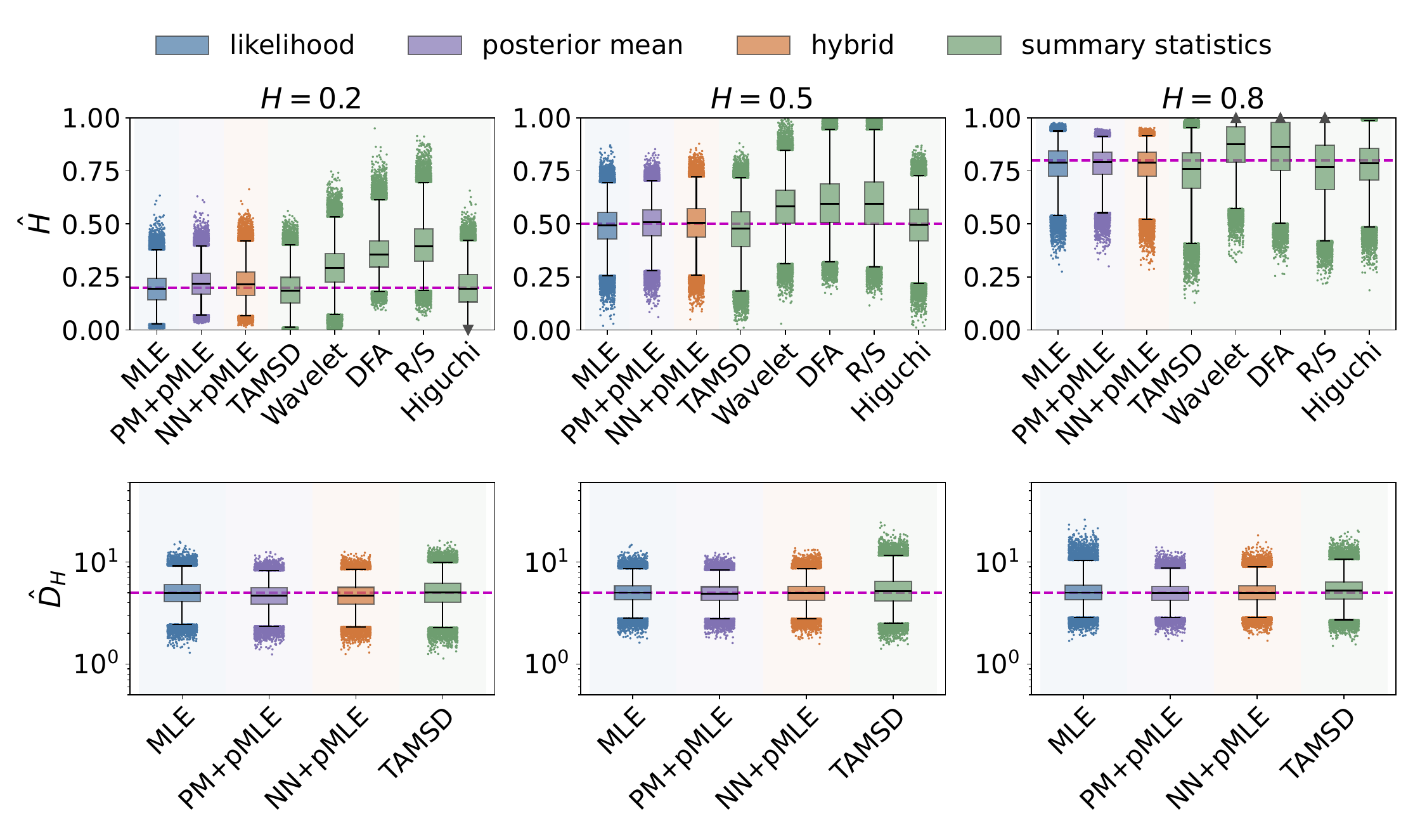}
    \caption{
        Plots of interquartile range (box), median (line), 1\textsuperscript{st} and 99\textsuperscript{th} percentiles (whiskers) and the tails (dots) for estimated $\hat{H}$ (top row) and $\hat{D}_H$ (bottom row) from $10^5$ fBM trajectories.
        The three columns show the same plots for $H=0.2$, $0.5$ and $0.8$.
        The magenta dashed line shows the true parameter values.
        }
    \label{fig:compare_estimators_clean}
\end{figure}

Table \ref{tab:estimator_comparison} shows for each estimator the root mean squared error (RMSE) in $\hat{H}$ and Figure \ref{fig:compare_estimators_clean} (top row) the full distributions.
A difference between two estimation methods is a paired quantity with a much smaller Monte-Carlo error compared to previous sections, since each estimation method is given the same set of trajectories.
The bold entries denote the lowest RMSE in each column of $H$ values, with PM+pMLE having the lowest RMSE for all columns.
Every other estimator is resolved as worse through a paired bootstrap (Section \ref{sec:estimator_implementation}).
The reason that PM+pMLE falls below the Cram\'er--Rao lower bound at $H=0.8$ is because it is a biased (shrinkage) estimator and the bound only applies to unbiased estimators.
As expected, the other summary statistics based approaches have higher RMSE and DFA and R/S are biased upwards at small $H$ due to finite series data \cite{mandelbrot1969robustness}.

\begin{figure}
    \centering
    \includegraphics[width=\linewidth]{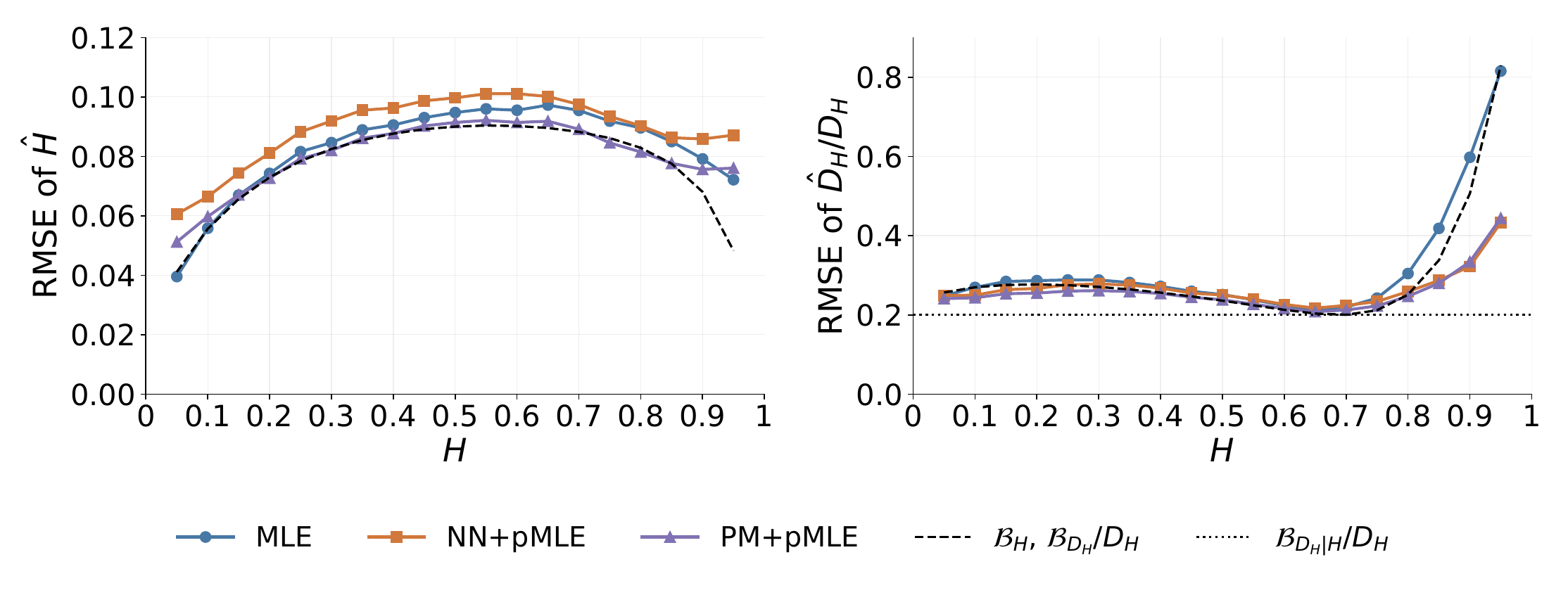}
    \caption{The root mean squared error (RMSE) of estimated $\hat{H}$ (\textit{left}) and RMSE of estimated $\hat{D}_H$ as a ratio of $D_H$ (\textit{right}) against $H$.
    The RMSE was calculated for $10^4$ fBM trajectories per data point with $K=50$, $t_K=100$, $\sigma_\epsilon=0$ and $D_H$ drawn log-uniformly on $[1.25,20]$ for every trajectory.
    Colours and markers denote the different estimation methods: MLE (blue circles), NN+pMLE (orange squares) and PM+pMLE (purple triangles).
    The dashed lines show Cram\'er-Rao bounds $\mathcal{B}_H$ and $\mathcal{B}_{D_H}/D_H$ for unbiased estimators and the dotted line is $\mathcal{B}_{D_H|H}/D_H = \sqrt{2/K}$, the bound on the amplitude with $H$ known \eqref{eq:crb_D}.}
    \label{fig:h_scan}
\end{figure}

\begin{table}
    \centering
    \begin{tabular}{l|ccc|ccc}
        \multirow{2}{*}{Estimator} & \multicolumn{3}{c|}{$\textrm{RMSE} \log_{10}\hat{D}_H$} & \multicolumn{3}{c}{$\textrm{RMSE} \hat{D}_H$} \\
            & $H{=}0.2$ & $0.5$ & $0.8$ & $H{=}0.2$ & $0.5$ & $0.8$ \\
        \hline

        MLE & 0.123 & 0.104 & 0.115 & 1.45 & 1.25 & 1.54 \\

        PM$+$pMLE & \textbf{0.122} & \textbf{0.102} & \textbf{0.103} & \textbf{1.29} & \textbf{1.18} & \textbf{1.24} \\
        NN$+$pMLE & 0.126 & 0.104 & 0.104 & 1.35 & 1.24 & 1.28 \\
        TAMSD & {0.138} & {0.144} & {0.128} & {1.67} & {1.97} & {1.73} \\
        \hline
        $\mathcal{B}_{D_H|H}$ & 0.087 & 0.087 & 0.087 & 1.00 & 1.00 & 1.00 \\
        $\mathcal{B}_{D_H}$ & 0.120 & 0.102 & 0.109 & 1.38 & 1.18 & 1.26 \\
    \end{tabular}
    \caption{RMSE of $\hat{D}_H$ for estimators on one set of $10^5$ simulated trajectories without localisation error.
    Both the RMSE of $\log_{10}\hat{D}_H$ (left) and $\hat{D}_H$ (right) are shown because comparing relative and absolute errors change yields different results.
    The true value is $D_H=5$.
    The bold entries show the lowest RMSE as in Table \ref{tab:estimator_comparison}.
    The last two rows are the Cram\'er-Rao bounds with $H$ known $\mathcal{B}_{D_H|H}$ and unknown $\mathcal{B}_{D_H}$ from \eqref{eq:crb_D}.
    }
    \label{tab:joint_amplitude}
\end{table}

Next, we will examine $D_H$ for the MLE, PM+pMLE, NN+pMLE and TAMSD, since the other methods do not estimate the generalised diffusion coefficient.
Table \ref{tab:joint_amplitude} shows for each estimator the RMSE in $\hat{D}_H$ and Figure \ref{fig:compare_estimators_clean} (bottom row) the full distributions.
Clearly, PM+pMLE has the lowest RMSE $\log_{10}\hat{D}_H$ and RMSE $\hat{D}_H$ across all values of $H$.
There are some computational aspects that should also be considered between different methods.
The MLE methods depend on optimisation algorithms that are typically computationally expensive every time an estimation is produced.
On the other hand, the NN+pMLE method requires initial network training but is otherwise orders of magnitude less expensive than MLE for estimation.
Training the network took $10$~s on a laptop processor for $10^5$ trajectories of length $50$.
Having said this, the neural network hybrid method requires retraining for different trajectory lengths.
The posterior mean, PM+pMLE, needs no training or optimisation.
The summary statistic methods, such as TAMSD, are computationally the cheapest and can be applied to any trajectory length but sacrifice estimation accuracy and precision.

Figure \ref{fig:h_scan} shows the RMSE of $\hat{H}$ and $\hat{D}_H/D_H$ for the MLE, PM+pMLE and NN+pMLE across all values of $H$, with $D_H$ drawn at random for every trajectory.
The plot shows a wider range of $H$ values than the Tables \ref{tab:estimator_comparison} and \ref{tab:joint_amplitude}.
The MLE stays close to the Cram\'er-Rao bounds and beats NN+pMLE and PM+pMLE for values of $H$ close to $0$ or $1$ because both NN+pMLE and PM+pMLE shrink towards the interior region of $H$ (see Section \ref{sec:posterior_mean}).
For $\hat{D}_H$, all estimators follow the Cram\'er-Rao bound but as $H\rightarrow1$ the bound grows steeply because the correlation $\rho$ in \eqref{eq:fisher_inverse} approaches one.
While the MLE follows the Cram\'er-Rao bound as $H\rightarrow 1$, the biased estimators, NN+pMLE and PM+pMLE, do much better because the profiled generalised diffusion coefficient \eqref{eq:profiled_amplitude_traj} is sensitive to overestimating $H$ near $H=1$.
So the bias towards $H=0.5$ keeps $\hat{H}$ away from $H=1$ allowing better estimation of $D_H$ with a relatively small increase in RMSE of $\hat{H}$.
This shows the advantage of using MLE or NN+pMLE and PM+pMLE for different $H$ values and with consideration for estimating $D_H$.

\subsection{Cram\'er-Rao bounds}
\label{sec:crb}
The Cram\'er-Rao bounds computed at $\sigma_\epsilon=0$ in Tables \ref{tab:estimator_comparison} and \ref{tab:joint_amplitude} are the smallest standard deviation that an unbiased estimator can have.
These bounds are derived from the Fisher information matrix $\mathcal{I}$. which for a Gaussian process with covariance $D_H\mathbf{\Sigma}_0(H)$ is
\begin{equation}
    \mathcal{I} = \frac{1}{2}\begin{pmatrix} K/D_H^2 & \mathrm{tr}\,\mathbf{A}/D_H \\ \mathrm{tr}\,\mathbf{A}/D_H & \mathrm{tr}\,\mathbf{A}^2 \end{pmatrix},
    \qquad
    \mathbf{A} = \mathbf{\Sigma}_0(H)^{-1}\,\frac{\partial\mathbf{\Sigma}_0(H)}{\partial H},
    \label{eq:fisher_traj}
\end{equation}
where rows and columns are ordered as $(D_H,H)$.
Let $\mathcal{B}_{\theta}$ denote the bound on the standard deviation of $\hat{\theta}$ where the other parameter $\phi$ is unknown and $\mathcal{B}_{\theta|\phi}$ when $\phi$ is known.
Then,
\begin{equation}
    \mathcal{B}_{\theta|\phi} = \frac{1}{\sqrt{[\mathcal{I}]_{\theta\theta}}},
    \qquad
    \mathcal{B}_{\theta} = \sqrt{[\mathcal{I}^{-1}]_{\theta\theta}} = \frac{\mathcal{B}_{\theta|\phi}}{\sqrt{1-\rho^2}},
    \qquad
    \rho = -\frac{[\mathcal{I}]_{D_HH}}{\sqrt{[\mathcal{I}]_{D_HD_H}\,[\mathcal{I}]_{HH}}} = -\frac{\mathrm{tr}\,\mathbf{A}}{\sqrt{K\,\mathrm{tr}\,\mathbf{A}^2}},
    \label{eq:fisher_inverse}
\end{equation}
where $\rho$ is the correlation between the errors in $D_H$ and $H$.

Using \eqref{eq:fisher_inverse}, we can calculate the Cram\'er-Rao bounds for the RMSE of the exponent
\begin{equation}
    \mathcal{B}_{H|D_H} = \sqrt{\frac{2}{\mathrm{tr}\,\mathbf{A}^2}},
    \qquad
    \mathcal{B}_{H} = \sqrt{\frac{2K}{K\,\mathrm{tr}\,\mathbf{A}^2-(\mathrm{tr}\,\mathbf{A})^2}}.
    \label{eq:crb_H}
\end{equation}
Both bounds do not depend on $D_H$ and explaining why Figure \ref{fig:traj_knobs} (top left) shows the precision of $\hat{H}$ not changing based on the value of $D_H$.
In the same way, we can calculate the Cram\'er-Rao bounds for the RMSE of the generalised diffusion coefficient
\begin{equation}
    \mathcal{B}_{D_H|H} = D_H\sqrt{\frac{2}{K}},
    \qquad
    \mathcal{B}_{D_H} = \frac{D_H\sqrt{2/K}}{\sqrt{1-\rho^2}}.
    \label{eq:crb_D}
\end{equation}
The closed form estimate \eqref{eq:profiled_amplitude_traj} evaluated at true $H$ is equivalent to $\mathcal{B}_{D_H|H}$ because the quadratic form $\mathbf{x}^T\mathbf{\Sigma}_0(H)^{-1}\mathbf{x}$ is $D_H$ times a chi-square variable with $K$ degrees of freedom, which has standard deviation $\sqrt{2K}$.
For the RMSE of $\log_{10}D_H$, we use the delta method \cite{doob1935limiting} to change from $D_H$ to $\log_{10}D_H$, which involves multiplying the Fisher information matrix by $(D_H\ln 10)^2$.
So each bound in \eqref{eq:crb_D} is divided by $D_H\ln 10$ to obtain the Cram\'er-Rao bounds for the RMSE of $\log_{10}D_H$ and the numerical values are shown in Table \ref{tab:joint_amplitude}

\subsection{The posterior mean and connection to the neural network}
\label{sec:posterior_mean}

For a given trajectory, MLE assigns a score of how probable the trajectory was observed under a model and its parameters, in this instance fBM with $H$ and $D_H$.
This probability of observation is then maximised to find the best set of $(H,D_H)$ for the trajectory, which is the essence behind MLE.
The posterior mean instead reports the score-weighted average of all possible $H$ for a given trajectory.
The generalised diffusion coefficient is then estimated as a profile MLE (pMLE) given the estimate for $H$ using \eqref{eq:profiled_amplitude_traj}.
We derive the posterior mean estimate, $\hat{H}_{PM}$, for $H$ by taking the average of $H$ over the likelihood.
Assuming $\sigma_{\epsilon}=0$ and a single trajectory with $K$ time points, \eqref{eq:unit_covariance} gives $\mathbf{\Sigma}=D_H\mathbf{\Sigma}_0(H)$ with $\det\mathbf{\Sigma}=D_H^K\det\mathbf{\Sigma}_0(H)$ and $\mathbf{\Sigma}^{-1}=\mathbf{\Sigma}_0(H)^{-1}/D_H$.
The loglikelihood \eqref{eq:fbm_traj_loglikelihood} becomes
\begin{equation}
    \ell(H,D_H) = -\frac{K}{2}\ln 2\pi - \frac{K}{2}\ln D_H - \frac{1}{2}\ln\det\mathbf{\Sigma}_0(H) - \frac{q(H)}{2D_H},
    \label{eq:loglik_separated}
\end{equation}
where $q(H) = \mathbf{x}^T\mathbf{\Sigma}_0(H)^{-1}\mathbf{x}$.
Given $H$, we know that the closed form estimate for the generalised diffusion coefficient is \eqref{eq:profiled_amplitude_traj}, $\hat D_H(H)=q(H)/K$.
Substituting this into \eqref{eq:loglik_separated}, we obtain
\begin{equation}
    \ell_p(H) = -\frac{K}{2}\ln 2\pi + \frac{K}{2}\ln K -\frac{K}{2}\ln q(H) - \frac{1}{2}\ln\det\mathbf{\Sigma}_0(H) - \frac{K}{2}.
    \label{eq:profile_explicit}
\end{equation}
Maximising $\ell_p(H)$ in \eqref{eq:profile_explicit} over $H$ gives the MLE $\hat H$, which is the same as maximising \eqref{eq:loglik_separated} over $H$ and $D_H$ jointly and the score of a value for $H$ is $e^{\ell_p(H)}$.
To calculate this score from the joint likelihood \eqref{eq:loglik_separated}, we start with the joint probability density of $(H,D_H)$ and Bayes' theorem, which states that the joint probability density given the data is the normalised product of the likelihood and prior densities
\begin{equation}
    p(H,D_H\mid\mathbf{x}) = \frac{e^{\ell(H,D_H)}\pi(H)\pi(D_H)}{\int\int e^{\ell(H',D')}\pi(H')\pi(D')dH'dD'}.
    \label{eq:joint_posterior}
\end{equation}
We assume the prior on $H$ to be $\pi(H) = 1/0.98$ uniform on $[0.01,0.99]$ and the prior on $D_H$ to be $\pi(D_H)=1/D_H$ which is uniform in $\ln D_H$ \cite{jeffreys1946invariant,bishop2006pattern}.
The distribution of $H$, or the score, follows by integrating \eqref{eq:joint_posterior} over $D_H$,
\begin{equation}
    p(H\mid\mathbf{x}) = C \left[\det \mathbf{\Sigma}_0(H) \right]^{-\frac{1}{2}} \left( \frac{2}{q(H)} \right)^{\frac{K}{2}},
    \label{eq:marginal_posterior}
\end{equation}
where $C = (2\pi)^{-\frac{K}{2}}\Gamma\left(\frac{K}{2}\right) \pi(H)/\int\!\!\int e^{\ell(H',D')}\,\pi(H')\,\pi(D')\,dH'\,dD'$.
Finally, the posterior mean is the expected value of $H$ under \eqref{eq:marginal_posterior}
\begin{equation}
    \hat H_{\text{PM}} = \int_{0.01}^{0.99} H\,p(H\mid\mathbf{x})\,dH,
    \label{eq:posterior_mean}
\end{equation}
which is the estimate that minimises the expected squared error over trajectories drawn under the prior \cite{berger1985statistical,kay1993fundamentals,lehmann1998theory}.
In practice, the integral in \eqref{eq:posterior_mean} is evaluated on a discretized grid of $H_k=0.01+\Delta H k$ values for $k = 0,\cdots,N_H$ where $\Delta H=(0.99-0.01)/N_H$.
Once $\hat{H}_{\text{PM}}$ is evaluated, the profile MLE for $D_H$ \eqref{eq:profiled_amplitude_traj} is used to estimate $\hat{D}_H$ given $\hat{H}_{\text{PM}}$.
Since it is an expected value, the posterior mean estimates shrink towards the interior of the $H$ region and almost never returns a value on a boundary.
This explains the reason it has smaller root mean squared errors than the MLE since the largest MLE errors are its excursions towards $H=0$ and $H=1$.
However, the price of this smaller RMSE is that posterior mean estimates have a bias towards $H=0.5$ and Bayesian inference of $H$ from single trajectories have been explored previously \cite{krog2018bayesian,thapa2018bayesian}.
The advantage of the posterior mean is that no optimiser, starting values and training is needed to estimate $H$.

The neural network estimation of $H$ is an approximation of the posterior mean estimation where unlimited training data and network capacity would recover \eqref{eq:posterior_mean} exactly.
We let $g_{\bs{w}}(\bs{x})$ be the function that represents the neural network where $\bs{x}$ is the input trajectory and $\bs{w}$ contains the weights of the network.
After training, the final weights are $\hat{\bs{w}}$ and the neural network gives an estimate of $H$ given inputs such that
\begin{equation}
    \hat{H}_{\text{NN}} = g_{\hat{\bs{w}}}(\bs{x}).
    \label{eq:NNestimate}
\end{equation}
Note that the input of position $\bs{x}$ is normalised by the root mean squared position, which is the same as setting $D_H=1$ (see Section \ref{sec:estimator_implementation}).
The neural network is trained on $N_{\text{train}}$ samples of $(\bs{x}_i,H_i)$ where $H_i\sim \pi_{\text{train}}(H)$ and $\bs{x}_i$ is a simulated fBM trajectory with Hurst exponent $H_i$.
Training adjusts the weights to minimise the mean squared error between $g_{\bs{w}}(\bs{x}_i)$ and $H_i$ such that
\begin{equation}
    \hat{\bs{w}} = \operatorname*{argmin}_{\bs{w}} \left\{ \frac{1}{N_{\text{train}}} \sum_{i=1}^{N_{\text{train}}} \left( g_{\bs{w}}(\bs{x}_i)-H_i \right)^2 \right\},
\end{equation}
and in the limit as $N_{\text{train}}\rightarrow\infty$,
\begin{equation}
    \hat{\bs{w}} = \operatorname*{argmin}_{\bs{w}} \left\{ \mathbb{E}\left[ \left( g_{\bs{w}}(\bs{x}) - H \right)^2 \right] \right\}.
    \label{eq:trainingexpectation}
\end{equation}
The probability that a given trajectory $\bs{x}$ has exponent $H$ is
\begin{equation}
    p(H\mid \bs{x}) = \frac{p(\bs{x}\mid H)\pi_{\text{train}}(H)}{\int p(\bs{x}\mid H')\pi_{\text{train}}(H')dH'}.
\end{equation}
The mean exponent is then
\begin{equation}
    m(\bs{x}) = \int H p(H \mid \bs{x}) dH = \hat{H}_{\text{PM}},
\end{equation}
where the last equality holds using \eqref{eq:posterior_mean} when $\pi_{\text{train}}(H)$ is assumed to be the same as the prior on $H$ and because normalising the trajectories removes the scaling of $D_H$ in the same way as integrating over $D_H$ using the prior $1/D_H$ in \eqref{eq:marginal_posterior}.
Now, we can rewrite the expectation in \eqref{eq:trainingexpectation} as
\begin{equation}
    \mathbb{E}\left[ \left( g_{\bs{w}}(\bs{x}) - H \right)^2 \right] = \mathbb{E}\left[ \left( g_{\bs{w}}(\bs{x}) - \hat{H}_{\text{PM}} \right)^2 \right] + \mathbb{E}\left[ \left( \hat{H}_{\text{PM}} - H \right)^2 \right],
    \label{eq:nn_pythagoras}%
\end{equation}
where the cross term is zero since the posterior mean is defined as \eqref{eq:posterior_mean}.
Since $\mathbb{E}\left[ \left( \hat{H}_{\text{PM}} - H \right)^2 \right]$ does not depend on the trainable weights $\bs{w}$ of the neural network, \eqref{eq:trainingexpectation} becomes
\begin{equation}
    \hat{\bs{w}} = \operatorname*{argmin}_{\bs{w}} \left\{ \mathbb{E}\left[ \left( g_{\bs{w}}(\bs{x}) - \hat{H}_{\text{PM}} \right)^2 \right] \right\}.
\end{equation}
This shows that the neural network estimation of $H$ is an approximation of the posterior mean estimation \cite{hornik1989multilayer,leshno1993multilayer,sainsburydale2024likelihood}.

\begin{figure}
    \centering
    \includegraphics[width=\linewidth]{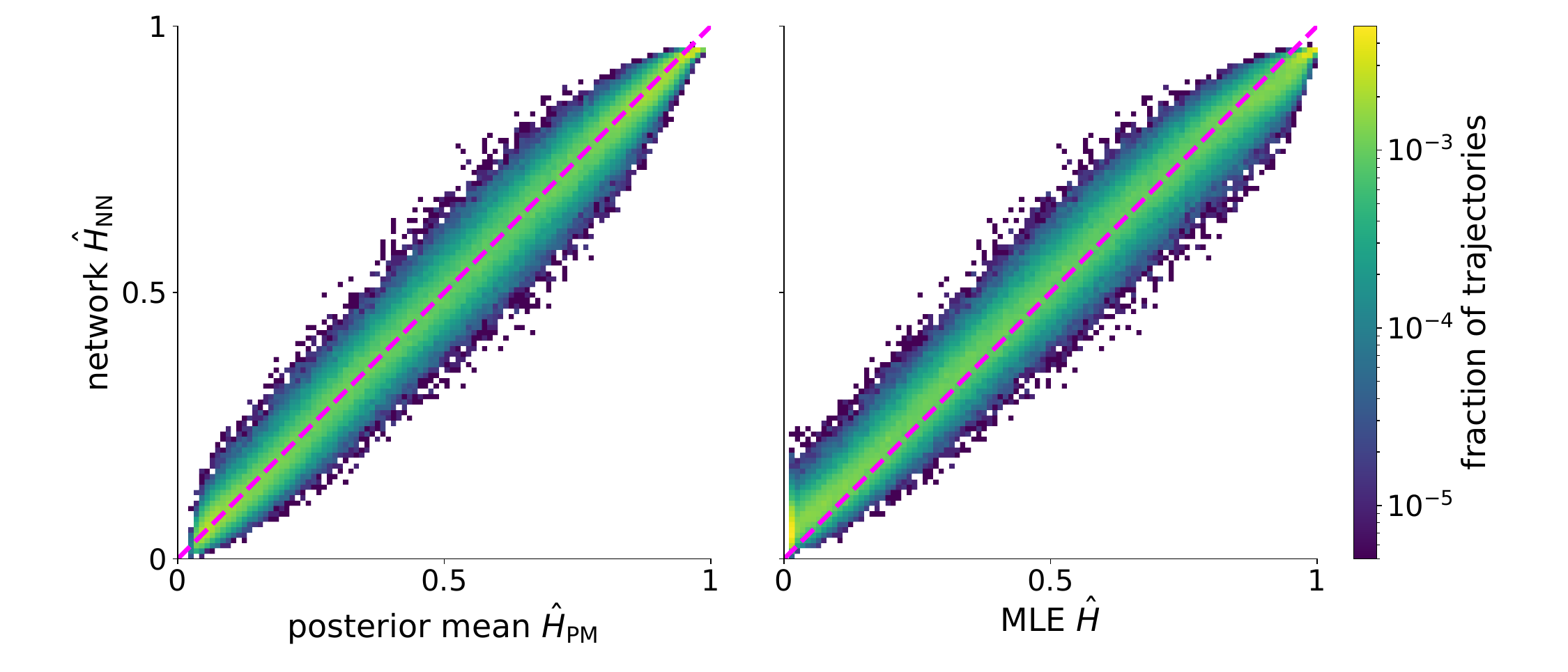}
    \caption{$\hat{H}_{\text{NN}}$ compared against $\hat{H}_{\text{PM}}$ \eqref{eq:posterior_mean} (\textit{left}) and against the MLE (\textit{right}).
    The network used in this paper was trained using $N_{\text{train}}=10^5$ and is described in Section \ref{sec:estimator_implementation}.
    The histograms used the same $2\times10^5$ fBM trajectories in Table~\ref{tab:nn_pm}.
    The colour denotes the fraction of the trajectories in each bin of width $0.01$ and the dashed magenta line marks equality.}
    \label{fig:nn_pm}
\end{figure}

To verify this numerically, we simulate $2\times10^5$ fBM trajectories with $H\sim\pi_{\text{train}}(H)$ and compare $\hat{H}_{\text{NN}}$ with $\hat{H}_{\text{PM}}$ for neural networks trained on $10^3$ to $10^8$ fBM trajectories in Table~\ref{tab:nn_pm}.
This clearly shows that $\hat{H}_{\text{NN}}$ and $\hat{H}_{\text{PM}}$ converge as $N_{\text{train}}$ increases and the consistency between the measured and predicted distance to $\hat{H}_{\text{PM}}$ verifies the theoretical prediction of \eqref{eq:nn_pythagoras}.
As $N_{\text{train}}$ is increased, the RMSE decreases and the approximation of $\hat{H}_{\text{NN}}$ becomes closer to $\hat{H}_{\text{PM}}$ as expected.
Figure~\ref{fig:nn_pm} shows a histogram of $\hat{H}_{\text{NN}}$ against $\hat{H}_{\text{PM}}$ (left) and MLE $\hat{H}$ (right).
It demonstrates that $\hat{H}_{\text{NN}}$ and $\hat{H}_{\text{PM}}$ have a linear relationship over all $H$ values with estimates centered on the magenta line, indicating the same bias structure.
However, $\hat{H}_{\text{NN}}$ underestimates $H$ compared to MLE as $H\rightarrow 1$ and overestimates as $H\rightarrow 0$, which is exactly the bias towards the interior that $\hat{H}_{\text{PM}}$ was expected to have.

\begin{table}
    \centering
    \begin{tabular}{lc|c|cc}
        \multirow{2}{*}{Nodes in the hidden layers} & \multirow{2}{*}{$N_{\mathrm{train}}$} & \multirow{2}{*}{RMSE} & \multicolumn{2}{c}{Distance to $\hat H_{\mathrm{PM}}$} \\
         & & & measured & predicted \\
        \hline
        $49$, $48$, $47$ & $10^3$ & 0.1624 & 0.1412 & 0.1411 \\
         & $10^4$ & 0.1106 & 0.0758 & 0.0761 \\
         & $10^5$ & 0.0892 & 0.0387 & 0.0388 \\
         & $10^6$ & 0.0832 & 0.0214 & 0.0216 \\
         & $10^8$ & 0.0825 & 0.0187 & 0.0187 \\
        $512$, $512$, $512$ & $10^8$ & 0.0808 & 0.0085 & 0.0086 \\
        \hline
    \end{tabular}
    \caption{Results from estimating $\hat{H}_{\text{NN}}$ and $\hat{H}_{\text{PM}}$ using the same $2\times 10^5$ fBM trajectories. $H$ was drawn uniformly from $[0.01,0.99]$, $K=50$ and $\sigma_\epsilon=0$. The left most column shows the number of nodes in each of the three layers of the neural network and $N_{\text{train}}$ is the number of fBM trajectories used for training. The root mean squared error (RMSE) between the estimated $\hat{H}_{\text{NN}}$ and the true $H$ is shown in the column titled RMSE. The measured distance to $\hat{H}_{\text{PM}}$ was calculated using the root mean squared difference between $\hat H_{\mathrm{NN}}$ and $\hat H_{\mathrm{PM}}$ over the $2\times 10^5$ test trajectories. The predicted distance to $\hat{H}_{\text{PM}}$ was calculated using $(\text{RMSE}^2-\text{RMSE}_{\text{PM}}^2)^{1/2}$ derived from \eqref{eq:nn_pythagoras} with $\text{RMSE}_{\text{PM}}=0.0803$. The posterior mean estimate was evaluated on an evenly spaced grid of $785$ values in $H$.}
    \label{tab:nn_pm}
\end{table}

\subsection{Application to experimental trajectories}
\label{sec:elife_application}

We now apply the parameter inference methods to intracellular vesicle transport data from \cite{han2020deciphering} and \cite{tirumala2024single}.
The data from \cite{han2020deciphering} are tracks of early endosomes (GFP-Rab5), sorting endosomes (GFP-SNX1) and lysosomes (LysoBrite) in MRC5 fibroblasts while the data from \cite{tirumala2024single} are EGF and Dextran cargoes in HeLa cells.
The datasets use different cells, tracking algorithms and microscopy equipment.
Each trajectory from \cite{han2020deciphering} and \cite{tirumala2024single} is cut into non-overlapping windows of $K+1=51$ datapoints and each window is translated such that every window trajectory begins at the origin.
Each window has $K=50$ non-zero positions in both $x$ and $y$ axes and since the first position of each window is subtracted from the others, the localisation error of that position is added to every datapoint within the window.
This means that the error term of the covariance \eqref{eq:unit_covariance} becomes $\sigma_\epsilon^2(\mathbf{I}+\mathbf{1}\mathbf{1}^{T})$.
Here, we set $\sigma_\epsilon$ by assuming that errors are independent between frames and using the difference between frames $x^o_j-x^o_{j-1} = x_j-x_{j-1} + \epsilon_j-\epsilon_{j-1}$ where $x^o_j$ is the observed co-ordinate, $x_j$ is the true co-ordinate and $\epsilon_j$ is the noise term at time indexed by $j$.
Since the mean squared displacement in one co-ordinate is $\mathbb{E}\left[(x^o_j-x^o_{j-1})^2\right]=\mathbb{E}\left[(x_j-x_{j-1})^2\right]+2\sigma_{\epsilon}^2$, we can write the single frame mean squared displacement for a window $w$ as
\begin{equation}
    \text{MSD}_w(1)=\mathbb{E}\big[(\Delta x_j)^2+(\Delta y_j)^2\big]+4\sigma_\epsilon^2.
    \label{eq:msd_window}
\end{equation}
We approximate $\text{MSD}_w(1)$ as the average over the $50$ steps in each window.
From \eqref{eq:msd_window}, the error is bounded from above for each window by
\begin{equation}
    \sigma_\epsilon^2\leq\frac{1}{4}\,\mathrm{MSD}_w(1),
    \label{eq:sigma_bound}
\end{equation}
where the equality only holds for a trajectory that does not move.
For every window of a video in the dataset, $\sigma_{\epsilon}$ is set to the $5$\textsuperscript{th} percentile value of the bound in \eqref{eq:sigma_bound}, over all the windows of that video.
We set $\sigma_{\epsilon}$ at this value since the already published experimental datasets \cite{han2020deciphering,tirumala2024single} do not provide calibration values.
The median of this approximated error across the dataset is $12$~nm for Rab5, $10$~nm for SNX1, $6$~nm for lysosomes, $21$~nm for EGF and $8$~nm for Dextran.

\begin{figure}
    \centering
    \includegraphics[width=0.95\linewidth]{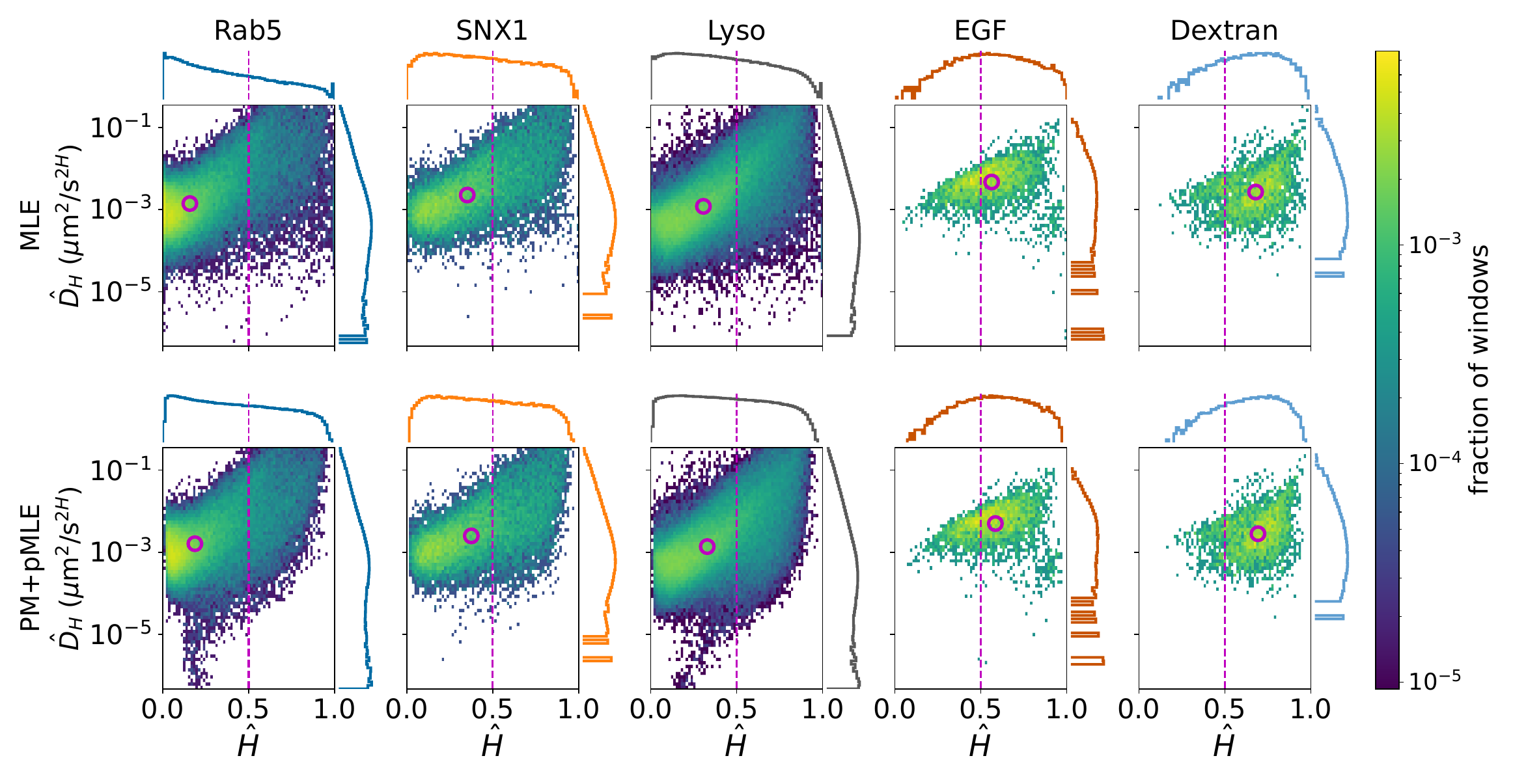}
    \caption{Joint distributions of $\hat H$ and $\hat D_H$ for intracellular vesicle datasets with localisation error set to the $5$\textsuperscript{th} percentile value of $\sigma_{\epsilon}$ over all windows for each video.
    The two-dimensional histograms for estimates using the MLE (\textit{top row}) and PM+pMLE (\textit{bottom row}) methods are shown with both marginals in $\hat H$ and $\hat D_H$ plotted above and to the right of each subplot.
    The magenta circle marks the median values and the magenta dashed line marks $H=0.5$.}
    \label{fig:elife_joint_HD_perwindow}
\end{figure}

Figure~\ref{fig:elife_joint_HD_perwindow} shows the estimated $\hat H$ and $\hat D_H$ values for each window across the intracellular vesicle datasets.
The two rows show parameter values estimated using MLE and PM+pMLE and their marginal distributions above and to the right of each subplot.
In agreement with \cite{han2020deciphering} and \cite{ba2018whole}, Figure~\ref{fig:elife_joint_HD_perwindow} demonstrates that early endosomes, sorting endosomes and lysosomes are largely subdiffusive ($H<0.5$).
Surprisingly, EGF and Dextran cargoes are mostly superdiffusive ($H>0.5$), although this could be due to selection of the most active cargo \cite{tirumala2024single}.
A striking visual finding is that there are no clusters of $\hat H$ or $\hat D_H$, which one would expect to find if there existed distinct modes of motion.
Instead, for each dataset, there is a smooth range across $\hat H$ or $\hat D_H$.
A limitation here is that the videos and trajectories may not have the spatial and/or temporal resolution to discern distinct modes of intracellular transport.
Further experiments will be needed to test if intracellular transport occurs in distinct modes or is a continuous spectrum of persistence and diffusivity.
Figure~\ref{fig:elife_joint_HD_perwindow} also shows that vesicular compartments (Rab5, SNX1 and lysosomes) move differently to cargoes (EGF and Dextran), while movement within types are largely similar.
One possible limitation here is that this could be due to two different experimental methods creating artefacts in the analysis.
As predicted in Section~\ref{sec:posterior_mean}, PM+pMLE has a bias towards $H=0.5$, which is evident when comparing the top and bottom row marginal distributions for $\hat H$.

Since each window has uncertainty in estimating $H$, we now quantify the uncertainty of classifying trajectories as subdiffusive or superdiffusive.
To measure the uncertainty, we use the posterior distribution of $H$ for each window, $p(H\mid\mathbf{x})$ in \eqref{eq:marginal_posterior}, whose mean \eqref{eq:posterior_mean} is $\hat H_{\text{PM}}$ in Figure~\ref{fig:elife_joint_HD_perwindow}.
Since $\sigma_\epsilon>0$, it is evaluated numerically on a grid described in Section~\ref{sec:estimator_implementation}.
The central $95\%$ credible interval of $p(H\mid\mathbf{x})$ has a median width of $0.25$ over the MRC5 windows and $0.34$ over the HeLa windows.
The probability that a window is subdiffusive is
\begin{equation}
    P_{\text{sub}}(\mathbf{x}) = \int_{0.01}^{0.5} p(H\mid\mathbf{x})\,dH,
    \label{eq:p_sub}
\end{equation}
and the probability that it is superdiffusive is $1-P_{\text{sub}}(\mathbf{x})$.
A window is classified as subdiffusive when $P_{\text{sub}}(\mathbf{x})>0.95$ and superdiffusive when $P_{\text{sub}}(\mathbf{x})<0.05$, and undetermined otherwise.
Using this classification rule, $60.0\%$ of windows from MRC5 trajectories \cite{han2020deciphering} are subdiffusive, $11.5\%$ superdiffusive and $28.5\%$ undetermined.
On the other hand, $4.6\%$ of windows from HeLa trajectories \cite{tirumala2024single} are subdiffusive, $47.7\%$ superdiffusive and $47.7\%$ undetermined.

A trajectory $n$ is cut into $W_n\geq2$ windows $\mathbf{x}_{n,1},\dots,\mathbf{x}_{n,W_n}$ and it is determined to exhibit both subdiffusion and superdiffusion if
\begin{equation}
    \min_{w=1,\dots,W_n} P_{\text{sub}}(\mathbf{x}_{n,w}) < 0.05 \quad\text{and}\quad \max_{w=1,\dots,W_n} P_{\text{sub}}(\mathbf{x}_{n,w}) > 0.95.
    \label{eq:both_types}
\end{equation}
Out of $17{,}727$ MRC5 trajectories with at least two windows, $26\%$ satisfy \eqref{eq:both_types} and out of $397$ HeLa trajectories, $21\%$ satisfy \eqref{eq:both_types}.
Due to the uncertainty in estimating $H$, a trajectory with constant $H$ could still satisfy \eqref{eq:both_types} by chance.
To test how likely a constant $H$ trajectory could be mistaken for a trajectory exhibiting both subdiffusion and superdiffusion, we simulated each trajectory again with fBM using a constant $H$ set to the median MLE $\hat H$ and $D_H$ set to the median MLE $\hat D_H$ of its $W_n$ windows.
The localisation error, $\sigma_\epsilon$, and frame time are kept the same.
Each simulated window is then classified as the experimental data was using \eqref{eq:p_sub}.
Only $2\%$ of the simulated MRC5 trajectories and $7\%$ of the simulated HeLa trajectories satisfy \eqref{eq:both_types}, showing that the experimental trajectories have significantly more trajectories exhibiting both subdiffusion and superdiffusion than by chance.
If each window is classified by the raw estimated $\hat H_{\text{PM}}$, $47\%$ of MRC5 and $67\%$ of HeLa trajectories contain windows that are both subdiffusive and superdiffusive compared with $27\%$ (MRC5) and $54\%$ (HeLa) for simulated trajectories with constant $H$.

These results demonstrate that intracellular cargoes move with different $H$ across times rather than being a heterogeneous population with fixed $H$.
Instead of classifying data into exclusively subdiffusion or superdiffusion, intracellular transport displays time-heterogeneous movement behaviour with both subdiffusion and superdiffusion occurring within the same trajectory.

\subsection{Implementation details}
\label{sec:estimator_implementation}

\textbf{Simulated trajectories.}
In Section \ref{sec:traj_accuracy_precision}, the fBM trajectories are generated from fractional Gaussian noise with the \texttt{fbm} package \cite{flynn2020fbm} using the Davies--Harte method \cite{davies1987tests,bilokon2026efficient}.
Section \ref{sec:estimator_comparison} uses the same method and each estimator fits the same set of trajectories making the Tables \ref{tab:estimator_comparison} and \ref{tab:joint_amplitude} a fair paired comparison.

\noindent
\textbf{Paired bootstrap.}
The bold in the Tables is decided by a paired bootstrap confidence interval for the difference between estimators \cite{efron1993introduction}.
Since every estimator is applied to the same set of $10^5$ trajectories, the difference between the RMSE of two estimators, $\Delta = \mathrm{RMSE}_a - \mathrm{RMSE}_b$, is a paired quantity.
The standard error of $\Delta$ can then be estimated by resampling.
For 2000 bootstrap replicates, we draw $10^5$ trajectory indices with replacement, recompute the RMSE and take their difference.
The standard deviation of the bootstrap replicates is the standard error $s_{\Delta}$.
In Tables \ref{tab:estimator_comparison} and \ref{tab:joint_amplitude}, the estimator with the lowest error is bold, and every other estimator whose difference from it has a $95\%$ confidence interval, $\Delta \pm 1.96\,s_{\Delta}$, that contains zero is also bold.
A column with more than one bold entry is a tie at the resolution of the simulation and not a rank.

\noindent
\textbf{Likelihood optimisation.}
The log-likelihood \eqref{eq:fbm_traj_loglikelihood} is computed from the Cholesky factorisation $\mathbf{\Sigma}=\mathbf{L}\mathbf{L}^T$, with $\ln\det\mathbf{\Sigma}=2\sum_i\ln L_{ii}$ and $\mathbf{x}^T\mathbf{\Sigma}^{-1}\mathbf{x}=\|\mathbf{L}^{-1}\mathbf{x}\|^2$ obtained by forward substitution, which avoids forming $\mathbf{\Sigma}^{-1}$ and stays accurate when $\mathbf{\Sigma}_0$ is ill-conditioned near $H=1$.
A small relative ridge of $10^{-10}$ times the mean diagonal is added to the diagonal of $\mathbf{\Sigma}_0$ so that rounding error cannot make the factorisation fail as $H\to1$.
The log-likelihood is maximised with L-BFGS-B \cite{byrd1995limited,virtanen2020scipy} on $(\log_{10}D_H,H)$ over $D_H\in[10^{-3},500]$ and $H\in[0.01,0.99]$, keeping the best result from four random starts uniform in $\log_{10}D_H$ and $H$.
Unless otherwise stated, the MLE assumes $\sigma_{\epsilon} = 0$.

\noindent
\textbf{Posterior mean with localisation error.}
For the windowed estimation of Section~\ref{sec:elife_application}, $\sigma_\epsilon>0$ so the covariance becomes $D_H\mathbf{\Sigma}_0(H)+\sigma_\epsilon^2(\mathbf{I}+\mathbf{1}\mathbf{1}^T)$.
In this case, $D_H$ no longer multiplies the whole covariance, so it cannot be integrated out in closed form to give \eqref{eq:marginal_posterior} and \eqref{eq:profiled_amplitude_traj} is no longer the best estimate for $D_H$.
Instead, we evaluate the loglikelihood of each window summed over a grid of $50$ values of $H$ evenly spaced on $[0.01,0.99]$ and $41$ values of $\log_{10}D_H$ spaced by $0.1$ over two decades either side of the window's amplitude \eqref{eq:profiled_amplitude_traj} using $H=0.5$.
By adding $e^{\ell}$ over the grid of $D_H$ yields $p(H\mid\mathbf{x})$ with the prior $\pi(D_H)=1/D_H$ and $\hat H_{\text{PM}}$ is \eqref{eq:posterior_mean} summed over the $50$ values of $H$.
The prior must be restricted when $\sigma_\epsilon>0$ because the likelihood can tend towards $D_H\to0$ and $1/D_H$ at $D_H=0$ does not converge.
The estimate $\hat D_H$ is the MLE over $D_H$ at the grid value nearest to $\hat H_{\text{PM}}$.

\noindent
\textbf{Neural network.}
The neural network architecture is the same as \cite{han2020deciphering} with 3 hidden layers of $n-l$ rectified-linear nodes for $l=1,2,3$ and one output node.
The input is the trajectory divided by its root mean squared displacement, $\mathbf{x}/s(\mathbf{x})$ with $s(\mathbf{x}) = \big(\tfrac{1}{n}\sum_i x(t_i)^2\big)^{1/2}$.
This normalisation allows the network to estimate $H$ for any $D_H$ since $H$ estimation is invariant under the input trajectory scaling $\mathbf{x}\mapsto c\mathbf{x}$.
The output node is logistic to ensure $\hat{H}\in(0,1)$.
The training set is $10^5$ exact fBM trajectories without localisation error with $H$ drawn uniformly from $(0.01,0.99)$ and $D_H=1$, with a further $5000$ trajectories for validation.
Training uses a mean-squared-error loss, AdamW \cite{loshchilov2019decoupled} with a cosine-annealed learning rate \cite{loshchilov2017sgdr} for $60$ epochs in PyTorch \cite{paszke2019pytorch}, keeping the checkpoint with the lowest validation error, a mean absolute error of $0.07$.

\noindent
\textbf{Summary statistics.}
The summary statistics follow the procedures of the cited papers:
the TAMSD as in \cite{kepten2015guidelines}, over lags $1$--$10$;
the wavelet variance as in \cite{percival2000wavelet}, from the Haar wavelet on the increments over levels $1$--$5$;
DFA as in \cite{kantelhardt2001detecting}, with linear detrending over boxes of $4$--$12$ increments;
rescaled range (R/S) as in \cite{mandelbrot1969robustness}, over windows of $8$--$24$ increments;
and the fractal dimension as in \cite{higuchi1988approach}, over $k=1$--$5$.

\section{Discussion}
\label{sec:discussion}

In this paper, we have developed MLE of the Hurst exponent $H$ and the generalised diffusion coefficient $D_H$ from fBM given two different data types, densities and trajectories, typically produced by experiments.
We have measured its accuracy and precision with repeated numerical simulations.
We then compare MLE with a hybrid neural network estimation method \cite{han2020deciphering}, the posterior mean of $H$ with profile MLE of $D_H$ and five summary statistics methods on identical simulated trajectories.
We then apply the MLE and the posterior mean methods to trajectories from different intracellular cargoes from two different cell types and experiments established in literature \cite{han2020deciphering,tirumala2024single}.

Density data can only identify fBM parameters through several observation times and faces a non-identifiability problem given only one observation time.
Given several observation times, the MLE bias is negligible and MLE precision follows the Fisher information \eqref{eq:FisherInfoPoisson}, improving fastest as the number of observation times increase (Figure~\ref{fig:optimisation_trials_collapse}).
Even so, density data cannot distinguish fBM from other underlying processes such as scaled Brownian motion \cite{wang2022anomalous,sposini2022towards}.

Trajectory data identifies both parameters and is able to distinguish fBM from other underlying processes since it contains the correlation between positions at different times \eqref{eq:fbm_covariance}.
The number of observation times is again the most important quantity for precision in inferring fBM parameters from trajectory data.
We also point out the inherent issues with the loglikelihood and identifiability in MLE for $H=0$ and $H=1$.
This leads to the recommendation that localisation error should be measured beforehand for each experimental system, for example, by tracking a control set of immobile particles.
We also present a method, PM+pMLE, using the posterior mean of $H$ paired with the profiled generalised diffusion coefficient \eqref{eq:profiled_amplitude_traj}, which has the lowest error at $H=0.2$, $0.5$ and $0.8$ (Tables~\ref{tab:estimator_comparison} and \ref{tab:joint_amplitude}) and most other values of $H$ (Figure~\ref{fig:h_scan}).
We show that this method uses bias towards $H=0.5$ to obtain a lower RMSE than MLE and explain how the neural network, NN+pMLE, based on \cite{han2020deciphering} is an approximation of the posterior mean.
The comparison between the posterior mean and a neural network trained to minimise squared error on a parameter inference task is in general known \cite{hornik1989multilayer,leshno1993multilayer,sainsburydale2024likelihood} and can be applied to harder inference tasks \cite{cranmer2020frontier,lenzi2023neural} with proper uncertainty quantification \cite{seckler2022bayesian}.
Section \ref{sec:posterior_mean} explicitly shows the relationship between the posterior mean and a neural network trained to estimate fBM parameters.
Table \ref{tab:nn_pm} shows the decreasing distance between $\hat{H}_{\text{PM}}$ and $\hat{H}_{\text{NN}}$ as the number of training trajectories increases.
The estimators also differ in computational cost since the MLE needs optimisation for each trajectory and the neural network needs training for each trajectory length but the posterior mean needs neither.
For fBM with a known localisation error, we recommend using PM+pMLE and only using MLE where an estimate without shrinkage is required or when $H$ is close to its bounds ($0$ or $1$).

When applied to experimental data, we found that MLE and PM+pMLE show largely consistent results and corroborate previous experimental observations for intracellular cargoes \cite{han2020deciphering,ba2018whole,tirumala2024single}.
Surprisingly, EGF or Dextran cargoes were predominantly superdiffusive (Figure~\ref{fig:elife_joint_HD_perwindow}), which could be an artefact of trajectory selection \cite{tirumala2024single}.
Within each experiment (MRC5 and HeLa), the joint distributions of $H$ and $D_H$ were similar across markers and the MLE and PM+pMLE gave largely the same distributions, which is to be expected for markers targeting similar intracellular vesicle pathways.
Across the datasets, there were no clusters of $H$ or $D_H$ giving no evidence of distinct modes of motion and single trajectories often possessed both subdiffusive and superdiffusive windows.
In Section \ref{sec:elife_application} we showed that $71.5\%$ of windows of $50$ time points from the MRC5 trajectories and $52.3\%$ of windows from HeLa trajectories can be classified as subdiffusive or superdiffusive with a posterior probability above $0.95$.
Then, we found that $26\%$ of MRC5 trajectories and $21\%$ of HeLa trajectories contained both subdiffusive and superdiffusive windows compared with $2\%$ and $7\%$ of simulated constant-$H$ trajectories, showing that the estimation uncertainty of $H$ cannot account for the heterogeneous transport that is observed in experimental data.
These results show that intracellular cargo exhibit subdiffusion and superdiffusion at different times, when viewed through the fBM model.
A major limitation that still exists is each window giving one inferred pair of parameters $(H,D_H)$, so changes of motion within a window is undetectable.
This limits testing of multifractional Brownian motion models \cite{ayache2005multifractional,balcerek2022fractional} and more advanced techniques \cite{korabel2025deep} should be investigated with uncertainty quantified.

\section*{Acknowledgements}
D.S.H. acknowledges funding from the University of New South Wales Faculty Research Grant. V.A. acknowledges funding from EMBL Australia. M.J.S acknowledges funding from the Australian Research Council (DP230100025).

\bibliography{mainbib}
\bibliographystyle{unsrt}

\end{document}